\documentclass[twocolumn,times,tighten,twocolappendix]{aastex631}

\newcommand{\ergs}{$\rm erg~s^{-1}$}
\newcommand{\ergscm}{$\rm erg~s^{-1}~cm^{-2}$}
\newcommand{\ergscma}{$\rm erg~s^{-1}~cm^{-2}~\AA^{-1}$}
\newcommand{\fc}{$F_{\rm 5100}$}
\newcommand{\feii}{Fe {\sc ii}}
\newcommand{\fhb}{$F_{\rm H\beta}$}
\newcommand{\fph}{$F_{\rm phot}$}
\newcommand{\fvar}{$F_{\rm var}$}
\newcommand{\heii}{He {\sc ii}}
\newcommand{\ha}{H$\alpha$}
\newcommand{\hb}{H$\beta$}
\newcommand{\hg}{H$\gamma$}
\newcommand{\hd}{H$\delta$}
\newcommand{\kms}{$\rm km~s^{-1}$}
\newcommand{\mbh}{$M_{\rm BH}$}
\newcommand{\oiii}{[O {\sc iii}]}
\newcommand{\rmax}{$r_{\rm max}$}
\newcommand{\rfe}{$R_{\rm Fe}$}
\newcommand{\rl}{$R_{\rm BLR}$--$L$}
\newcommand{\sline}{$\sigma_{\rm line}$}
\newcommand{\tph}{$\tau_{\rm ph}$}
\newcommand{\tsp}{$\tau_{\rm sp}$}

\newcommand{\ihep}{Key Laboratory for Particle Astrophysics, Institute of High
Energy Physics, Chinese Academy of Sciences, 19B Yuquan Road, Beijing 100049,
China; huc@ihep.ac.cn, wangjm@ihep.ac.cn}
\newcommand{\ucas}{School of Physics, University of Chinese Academy of
Sciences, Beijing 100049, China}
\newcommand{\naoc}{National Astronomical Observatories of China, The Chinese
Academy of Sciences, 20A Datun Road, Beijing 100020, China}

\shorttitle{Reverberation for 11 PG Quasars}
\shortauthors{Hu et al.}

\begin{document}

\title{Supermassive Black Holes with High Accretion Rates in Active Galactic
Nuclei. XIV. \\ Long-Duration High-Cadence Reverberation Mapping Results for
11 PG Quasars}

\author{Chen Hu}
\affil{\ihep}

\author{Zhu-Heng Yao}
\affil{\ihep}
\affil{\ucas}

\author{Yong-Jie Chen}
\affil{\ihep}
\affil{Dongguan Neutron Science Center, 1 Zhongziyuan Road, Dongguan 523808,
China}

\author{Yu-Yang Songsheng}
\affil{\ihep}

\author{Yi-Lin Wang}
\affil{\ihep}
\affil{\ucas}

\author{Sen Yang}
\affil{\ihep}
\affil{\ucas}

\author{Hao Zhang}
\affil{\ihep}
\affil{\ucas}

\author{Wei-Jian Guo}
\affil{\naoc}

\author{Pu Du}
\affil{\ihep}

\author{Yan-Rong Li}
\affil{\ihep}

\author{Ming Xiao}
\affil{\ihep}

\author{Jun-Rong Liu}
\affil{\ihep}

\author{Hua-Rui Bai}
\affil{\ihep}
\affil{\ucas}

\author{Feng-Na Fang}
\affil{\ihep}
\affil{\ucas}

\author{Yi-Xin Fu}
\affil{\ihep}
\affil{\ucas}

\author{Yue-Chang Peng}
\affil{\ihep}
\affil{\ucas}

\author{Shuo Zhai}
\affil{\naoc}

\author{Jin-Ming Bai}
\affil{Yunnan Observatories, The Chinese Academy of Sciences, Kunming 650011,
China}

\author{Luis C. Ho}
\affil{Kavli Institute for Astronomy and Astrophysics, Peking University,
Beijing 100871, China}
\affil{Department of Astronomy, School of Physics, Peking University, Beijing
100871, China}

\author{Michael S. Brotherton}
\affil{Department of Physics and Astronomy, University of Wyoming, Laramie, WY
82071, USA}

\author{Jes\'us Aceituno}
\affil{Centro Astron\'omico Hispano en Andaluc\'ia, Sierra de los filabres sn,
E-04550 Gergal, Almer\'ia, Spain}
\affil{Instituto de Astrof\'isica de Andaluc\'ia (CSIC), Glorieta de la
astronom\'ia sn, E-18008 Granada, Spain}

\author{Hartmut Winkler}
\affil{Department of Physics, University of Johannesburg, PO Box 524, 2006
Auckland Park, South Africa}

\author{Jian-Min Wang}
\affil{\ihep}
\affil{School of Astronomy and Space Science, University of Chinese Academy of
Sciences, Beijing 100049, China}
\affil{\naoc}

\collaboration{23}{(SEAMBH collaboration)}

\begin{abstract}
  We report the results of a long-duration high-cadence reverberation mapping
  campaign of a second batch of 11 PG quasars using the 2.2m telescope at the
  Calar Alto Observatory. This follows a similar earlier study of another
  sample of 15 objects reported by \citet{hu21}. Among the 11 PG quasars, 8
  objects have the \hb\ time lags measured for the first time, while the other
  3 objects were observed in previous campaigns, but only had highly uncertain
  \hb-lag measurements. Long-term light curves are presented of photometric
  $V$-band, spectroscopic 5100 \AA\ continuum, and the \hb\ emission line,
  lasting for $\sim$3--6 years with a cadence of $\sim$6--14 days. Accurate
  \hb\ time lags ranging from $\sim$20 to 150 days in the rest frame are
  obtained. The estimated virial masses of the central supermassive black
  holes range from $\sim$(3--300)$\times10^7 M_\odot$. Combining these
  results with those reported in \citet{hu21}, we now have 26 PG quasars,
  with representative properties, having reliable \hb\ time-lag measurements
  from our long-duration high-cadence campaign. A tentative fit to the
  relation between the \hb\ time lag and the continuum luminosity for these 26
  objects gives a slope of 0.53.
\end{abstract}


\section{Introduction}
\label{sec-intro}

Reverberation mapping \citep{blandford82} has been used for several decades to
determine the size of the broad-line region (BLR) of active galactic nuclei
(AGNs) through measuring the time delay between the continuum light curve and
the response of the broad emission lines. Combining reverberation mapping with
direct angular size measurements of the BLR through optical interferometry
allows the measurement of the cosmological distance by the method of
Spectroastrometry and Reverberation Mapping (SARM), as was recently shown by
\citealt{wang20,li22,li23}. By itself, reverberation mapping is typically
used to determine the mass of the super massive black hole in the center of
an AGN \citep[e.g.,][]{peterson04}. Reverberation mapping has also established
the relation between the size of the BLR and the luminosity of the AGN
continuum (the \rl\ relation; e.g., \citealt{kaspi00,bentz13,du19}), which is
widely applied to estimate black hole masses using only single-epoch spectra
\citep[e.g.,][]{vestergaard02}.

\hb-lag measurements are reported for $\sim$250 AGNs in the literature,
including traditional single-object spectroscopic campaigns
\citep[e.g.,][]{peterson98a,kaspi00,bentz09,denney10,grier12,du14,barth15,fausnaugh17,du18b,hu21,bao22,u22,woo24,zastrocky24},
and multi-object spectroscopic surveys \citep[e.g.][]{grier17,malik23,shen24}.
Although the number of objects has greatly increased recently, the pool of AGN
with successful reverberation mapping measurements is highly heterogeneous in
terms of 1) the properties of the objects, e.g., the luminosity and the
accretion rate, and 2) the observation and data quality, e.g., the flux
calibration and the sampling period (including both campaign duration and
cadence). The majority of these efforts assume that the narrow \oiii\ emission
line luminosity is invariant during the campaign and can be used for flux
calibration, biasing their selection toward objects with strong \oiii. For
the multi-object spectroscopic surveys, due to the fixed uniform campaign
duration and observing cadence for all the objects, response delays in AGN
with short time lag tend to be detected more easily, which leads to these
results being cataloged more frequently. In addition, the measured time lag
tends to be biased to a longer value in undersampled observation sequences
compared to the corresponding values determined from high-cadence campaigns
(see \citealt{hu20b}; \citealt{hu21} and references therein for examples).
Thus, some previous \hb\ lag measurements could be overestimated, especially
for AGNs with relatively high luminosities which are often monitored with low
cadences. Finally, due to the limited precision of the flux calibration, the
time lags of objects with large variability amplitudes are easier to measure,
another potential bias likely associated with the accretion rate
\citep{wilhite08}. Thus, it is valuable to monitor a complete sample, and to
do so with homogeneous observational settings.

In 2017, the SEAMBH (super-Eddington accreting massive black hole;
\citealt{du14}) and the MAHA (monitoring AGNs with \hb\ asymmetry;
\citealt{du18a}) collaborations began a long-duration reverberation mapping
campaign at the Calar Alto Observatory (CAHA; the Centro Astron\'omico
Hispanoen Andaluc\'ia), the Wyoming Infrared Observatory (WIRO), and also the
Yunnan Observatory, that spectroscopically monitors as many as possible of the
low-redshift PG quasar sample \citep{boroson92}. This sample is incomplete but
likely representative in many fundamental issues \citep[see,
e.g.,][]{wisotzki00,jester05}. The campaign is still in progress, and aims to
obtain precise, high-cadence light curves, not only to determine reliable time
lag measurements, but also to study the structures of the BLRs by recovering
the information of the transfer functions. 

The reverberation mapping results of several objects monitored at CAHA have
already been published, including PG 2130+099 \citep{hu20b,yao24}, PG 0026+129
\citep{hu20a}, a batch of 15 PG quasars \citep{hu21}, PG 1119+120
\citep{donnan23}, and also five SARM targets \citep{li24}. These studies
illustrate the point made earlier that long duration and high cadence sampling
are essential, not only for AGNs with high luminosities (and thus long time
lags; see \citealt{hu21} for more discussion), but also for those with
peculiar reverberation properties, including discrete multiple lags (e.g., PG
0026+129; \citealt{hu20a}), large lag changes between seasons (e.g., PG
2130+099; \citealt{hu20b}) and other abnormal phenomena (e.g., different
long-term trends between the continuum and emission-line light curves, or
holidays when lines fail to follow the continuum). Another reason for
requiring high-cadence and long-duration campaigns is the so called
``geometric dilution'' effect \citep{goad14}, which causes the measured time
lag to be affected by the time scale of the continuum variability. This effect
has been observed over a six-year period in PG 2130+099 by \citet{yao24}. All
these findings rely on high-cadence monitoring lasting for multiple years.

As of July 2024, we have completed monitoring a second batch of 11 PG quasars
at CAHA. Here we report their \hb\ time-lag measurements. We follow a similar
analysis as previously \citet{hu21}, giving the light curves (Section
\ref{sec-lc}) determined using the integration method, the \hb\ time-lag
(Section \ref{sec-ts}) and velocity-width (Section \ref{sec-width})
measurements, and the estimated black hole masses (Section \ref{sec-mass}).
The properties of the sample, the observations and data reductions are
presented in Section \ref{sec-samp} and \ref{sec-obs}, respectively. In
Section \ref{sec-dis}, the results of three objects are compared with those
from previous campaigns. Finally, the current CAHA PG quasars reverberation
mapping sample including the 11 objects here and the 15 objects from
\citet{hu21} are discussed collectively. More analysis of the combined data
set of these 26 targets, including spectral decomposition for lags of \heii\
and \feii\ emission lines, velocity-resolved delays, etc., will be presented
in forthcoming papers.

\section{Sample}
\label{sec-samp}

In coordination between the SEAMBH and the MAHA collaborations, the
spectroscopic monitoring of all the 87 quasars in \citet{boroson92} are
performed at CAHA, WIRO, and the Yunnan Observatory. Observations are
scheduled at the three observatories based on the objects' spectral
properties, observability, and the available telescope time. At CAHA, we give
priority to targets with the highest estimated dimensionless accretion rates.
Such objects tend to have weak \oiii\ intensities, suitable for the CAHA
campaign in which a comparison star is used for the flux calibration rather
than the weak narrow line (see Section \ref{sec-obs} below). The details of
the sample selection and observational scheduling of our targets have been
presented in \citet{hu21}. In summary, 49 PG quasars are planned to be
monitored at CAHA. The observations of these objects began gradually due to
the limitation of the telescope time, and we kept monitoring an object until
the reverberations of its emission lines were revealed without a doubt. By
July 2020 enough data had been accumulated to secure the measurements of \hb\
time lags for the first batch of 15 objects, and the results thereof were
published in \citet{hu21}. In 2024 July we completed the data collection for
the second batch of 11 objects, and this is covered in this work.
Including the three singly published PG quasars mentioned in Section
\ref{sec-intro} above, we have successfully measured time lags for 29 objects,
among the 40 objects that have been monitored for at least one observational
season. The observations of the last 9 objects were started just after the
completion of the second batch presented in this paper.

Table \ref{tab-obs} lists the names, redshifts ($z$) and V-band Galactic
extinctions ($A_V$) of these second-batch AGN, as well as time-sampling
details of the observations.  Note that $z$ in column (3) is defined by the
narrow \oiii\ $\lambda\lambda$4959,5007 lines in our mean spectrum, while
$A_V$ in column (4) is from the NASA/IPAC Extragalactic Database,%
\footnote{\url{https://ned.ipac.caltech.edu/}}
which is based on \citet{schlafly11}. Where applicable, column (2) also lists
an alternative name often used in the literature.

\begin{deluxetable*}{llccrrrr@{~~--~~}r}
  \tablewidth{0pt}
  \tablecaption{Objects and Observations
  \label{tab-obs}}
  \tablehead{
  \colhead{Object} & \colhead{Other Name} &
  \colhead{$z$} & \colhead{$A_V$} & 
  \colhead{$N_{\rm obs}$} & \colhead{$T_{\rm median}$} &
  \colhead{Duration} & \multicolumn{2}{c}{Begin and End Dates}
  \\
  \colhead{} & \colhead{} & \colhead{} & 
  \colhead{(mag)} & \colhead{} & \colhead{(days)} &
  \colhead{(days)} & \multicolumn{2}{c}{}
  \\
  \colhead{(1)} & \colhead{(2)} & \colhead{(3)} & \colhead{(4)} &
  \colhead{(5)} & \colhead{(6)} & \colhead{(7)} & \multicolumn{2}{c}{(8)}
  } 
  \startdata
PG 0157+001 &   Mrk 1014 & 0.1620 & 0.079 &  164 &  6 &  2018 & Jul 2017 & Feb 2023 \\
PG 0844+349 &    TON 951 & 0.0643 & 0.101 &  153 &  6 &  1447 & Oct 2018 & Sep 2022 \\
PG 1116+215 &   TON 1388 & 0.1754 & 0.062 &   96 & 10 &  1625 & Jan 2020 & Jul 2024 \\
PG 1121+422 &            & 0.2248 & 0.062 &   41 & 14 &   899 & Feb 2020 & Jul 2022 \\
PG 1229+204 &    Mrk 771 & 0.0638 & 0.074 &   69 &  8 &   906 & Feb 2020 & Jul 2022 \\
PG 1341+258 &    TON 730 & 0.0866 & 0.031 &   60 &  9 &   926 & Feb 2020 & Aug 2022 \\
PG 1352+183 &            & 0.1507 & 0.051 &   75 &  7 &   914 & Feb 2020 & Aug 2022 \\
PG 1411+442 &            & 0.0895 & 0.026 &  155 &  8 &  1596 & May 2017 & Sep 2021 \\
PG 1444+407 &            & 0.2663 & 0.038 &  121 & 10 &  1950 & May 2017 & Sep 2022 \\
PG 2233+134 &            & 0.3252 & 0.181 &  171 &  6 &  2002 & Jul 2017 & Jan 2023 \\
PG 2308+098 &   4C 09.72 & 0.4329 & 0.116 &   84 &  8 &  1195 & Sep 2020 & Jan 2024
  \enddata
  \tablecomments{
  Columns (2)--(4) list the alternative name, redshift, and $V$-band Galactic
  extinction for each of our objects, respectively. Columns (5)--(7) list the
  time-sampling details of our observations: number of spectroscopic epochs,
  median time interval, and the duration of our spectroscopic monitoring.
  Column (8) lists the start and end dates of our observations for each
  object.
  }
\end{deluxetable*}

\begin{figure}
  \centering
  \includegraphics[width=0.45\textwidth]{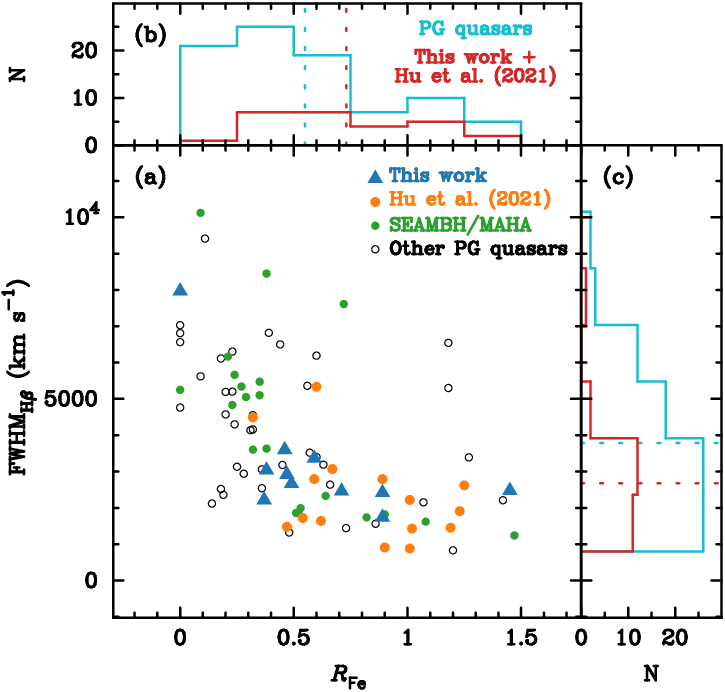}
  \caption{
  (a) The distribution of the objects in this work (blue triangles) and
  \citet{hu21} (orange solid cycles) in the plane of \hb\ FWHM versus the
  flux ratio of \feii\ to \hb\ (\rfe). Objects with published lags from other
  SEAMBH or MAHA observations (green solid cycles; see the text for the
  references), and other PG quasars in \citet{boroson92} (black empty cycles)
  are also plotted. Both measurements of \hb\ FWHM and \rfe\ are taken from
  \citet{boroson92} directly. (b) and (c) The histograms of \rfe\ and
  \hb\ FWHM for the CAHA sample (red) and the entire PG quasar sample (cyan),
  respectively. The mean values are indicated by the dotted lines.
  }
  \label{fig-ev}
\end{figure}

Panel (a) of Figure \ref{fig-ev} shows the distribution of the 11
objects presented in this work (blue triangles) and the 15 objects from
\citet{hu21} (orange solid cycles) in the so-called Eigenvector 1 plane
\citep[e.g.,][]{boroson92,shen14}, which is the full width at half-maximum
(FWHM) of the broad \hb\ emission line versus the relative intensity of \feii\
to \hb\ (\rfe).%
\footnote{
Note that, in order to maintain a consistent comparison of all PG quasars, the
values of \hb\ FWHM and \rfe\ plotted in Figure \ref{fig-ev} are always taken
from \citet{boroson92} even when these were remeasured in our campaign.
}
For comparison, 21 objects with published time-lag measurements from other
SEAMBH and MAHA observations
\citep{du14,huang19,hu20a,hu20b,bao22,donnan23,zastrocky24,yao24} are plotted
as green solid circles, and the other 41 objects in the \citet{boroson92}
sample are plotted as black open circles. The CAHA samples (the 15
objects in \citealt{hu21} plus the 11 objects in this work) are mainly
located at the high-accretion-rate end of EV1, in line with our chosen
selection criteria. Panels (b) and (c) show the histograms of \rfe\ and
\hb\ FWHM for the CAHA sample (red) and the entire PG quasar sample (cyan),
respectively. As marked by the dotted lines, the CAHA sample has $\sim$1/3
higher \rfe\ (0.73 versus 0.55) and $\sim$30\% narrower \hb\ FWHM (2677 versus
3785 \kms) on average compared to the entire PG quasar sample.

\begin{figure}
  \centering
  \includegraphics[width=0.45\textwidth]{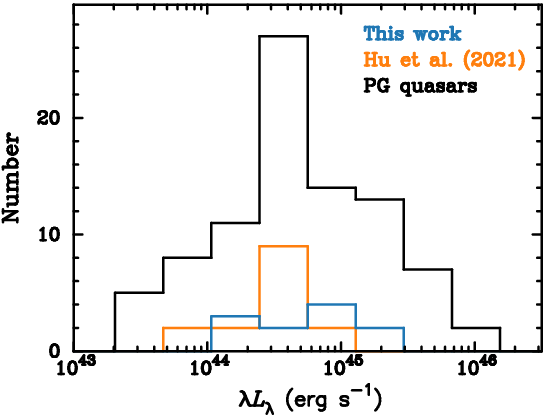}
  \caption{
  The histograms of the objects in this work (blue), \citet{hu21} (orange) and
  the entire PG quasar sample of \citet{boroson92} (black). The luminosities
  are calculated from the absolute magnitudes given in \citet{boroson92}.
  }
  \label{fig-lumi}
\end{figure}

Figure \ref{fig-lumi} shows histograms of the optical luminosity for
the objects in this work (blue), \citet{hu21} (orange), and the 87 PG quasars
of \citet{boroson92} with $z < 0.5$ (black).%
\footnote{
Note that the luminosities plotted in Figure \ref{fig-lumi} are all converted
from the absolute magnitudes at $\lambda$5500 in \citet{boroson92} to ensure
that these quantities were determined in a consistent manner. The same
cosmological parameters listed in Section \ref{sec-mass} were used for the
luminosities converted.
}
Due to the limited observation span available at the time, the \citet{hu21}
sample tends to have low-luminosity objects with lags ranging from
$\sim$20--100 days. Due to the longer monitoring, the sample in this work has
more high-luminosity objects of around $\sim10^{45}$ \ergs, which is important
for studying the \rl\ relation.

\section{Observations and Data Reduction}
\label{sec-obs}

The observations of four objects in this sample, PG 0157+001, PG 1411+442, PG
1444+407, and PG 2233+134, began in 2017 (see Table \ref{tab-obs}, col. 8).
They were among the targets that constituted the original CAHA campaign
sample, but for which further observations were required after three years, as
these had no reliable measurement of the \hb\ lag due to their slow or weak
variabilities. Results for the remainder of the original sample were published
in \citet{hu21}. The other seven objects in this paper were added to our
campaign later.

Columns (5), (6), and (7) of Table \ref{tab-obs} show the number of epochs,
median time sampling, and the $\sim$3--6 year time span (in days) of the
spectroscopic observations of each object, respectively. Our observations have
a cadence of $\sim$5--10 days for all of the targets except PG 1121+422.
Despite this, the data for PG 1121+422 is good enough for the \hb-lag
measurements in view of its large variability amplitude (see Figure
\ref{fig-lc1121} and Table \ref{tab-var} in the following sections).

The observing procedure, instrument settings, data reduction, and flux
calibration for the 11 objects in this work are exactly the same as for those
reported in \citet{hu20a,hu20b,hu21} and recorded there. Following are some
brief essentials.

The spectra were taken by the Calar Alto Faint Object Spectrograph (CAFOS) on
the CAHA 2.2m telescope, with Grism G-200 and a long slit set to a
3$\farcs$0-wide projection on the sky. The slit was rotated to enable the
simultaneous recording of the spectrum of a nearby unvarying comparison star.
Two exposures of spectra were taken per object for each epoch, with an
exposure time that typically yields the signal-to-noise ratio (S/N) of a
single spectrum better than $\sim$50 per pixel for the continuum around the
rest-frame wavelength 5100 \AA. Before taking the spectra, three broad-band
Johnson $V$-filter images were taken by CAFOS in the imaging mode, which also
served as acquisition camera for rotating the slit and aligning the target and
comparison star at the slit center.

The data reduction of the photometric and spectroscopic data were performed
using IRAF%
\footnote{
IRAF is distributed by the National Optical Astronomy Observatories, which are
operated by the Association of Universities for Research in Astronomy, Inc.,
under cooperative agreement with the National Science Foundation.
}
following standard procedures. The same 10$\farcs$6$\times$3$\farcs$0 spectral
extraction aperture was used for all objects. The reduced spectra cover the
wavelength range of $\sim$4000--8000 \AA\ with a dispersion of 4.47 \AA\
pixel$^{-1}$ and an average instrumental broadening of $\sim$1000 \kms\ in
FWHM. Each spectrum was flux-calibrated to an accuracy better than $\sim$3\%
using the sensitivity function determined by the simultaneously observed
spectrum of the comparison star (see \citealt{hu20a,hu20b} for details). 

\begin{figure*}
  \centering
  \includegraphics[width=0.88\textwidth]{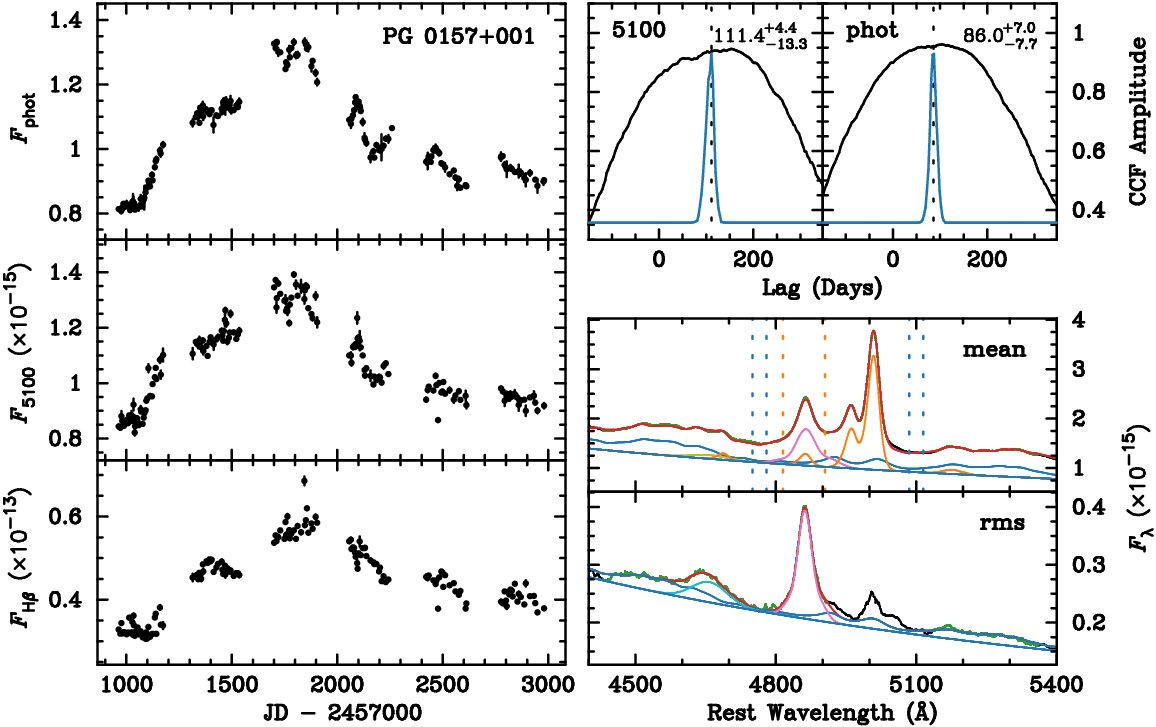}
  \caption{
  Light curves, ICCF analysis, and spectral decomposition of the mean and rms
  spectra for PG 0157+001. Left column: CAHA light curves of the photometric
  $V$-band flux (\fph) in arbitrary linear units (top panel), the
  spectroscopic continuum flux at 5100 \AA\ (\fc) in units of $10^{-15}$
  \ergscma\ (middle panel), and the integrated \hb\ flux (\fhb) in units of
  $10^{-13}$ \ergscm\ (bottom panel). Right column, two top panels: the CCFs
  (in black) and the corresponding CCCDs (in blue) for \hb\ with respect to
  \fc\ (adopted) and \fph, respectively. The time lag (in the observed frame)
  is marked by the vertically dotted line and displayed as the number with the
  errors in each panel. Right column, two lower panels: the mean and rms
  spectra after Galactic extinction correction (in green and black for parts
  in and out of the fitting windows, respectively) and the best-fit models (in
  red) comprised by spectral components including the AGN power-law continuum
  (in blue), \feii\ (in blue), broad \hb\ (in magenta), broad \heii\ (in
  cyan), narrow lines (in orange, only in the mean spectrum), and the host
  starlight (low and out of the panel window). In the panel of the mean
  spectra, the integration windows for the continuum and the \hb\ line are
  also marked by the vertical dotted lines in blue and orange, respectively.
  }
  \label{fig-lc0157}
\end{figure*}

\begin{figure*}
  \centering
  \includegraphics[width=0.88\textwidth]{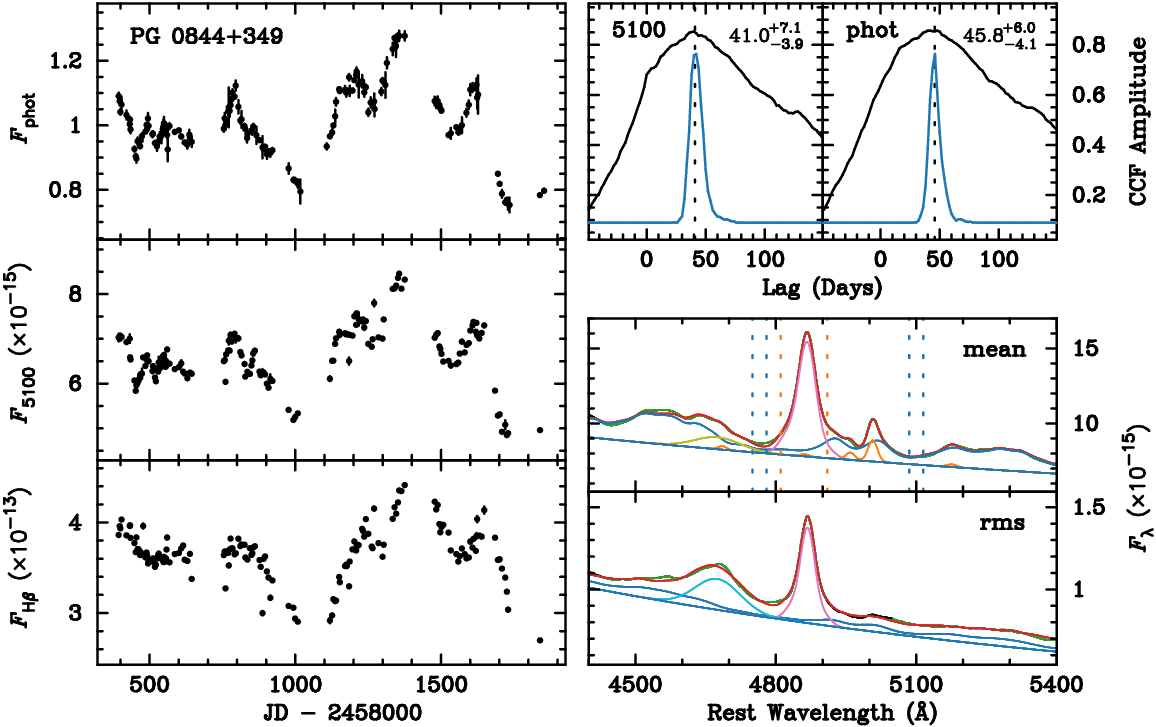}
  \caption{
  Light curves, ICCF analysis, and spectral decomposition of the mean and rms
  spectra for PG 0844+349, plotted in the same manner as in Figure
  \ref{fig-lc0157}.
  }
  \label{fig-lc0844}
\end{figure*}

\begin{figure*}
  \centering
  \includegraphics[width=0.88\textwidth]{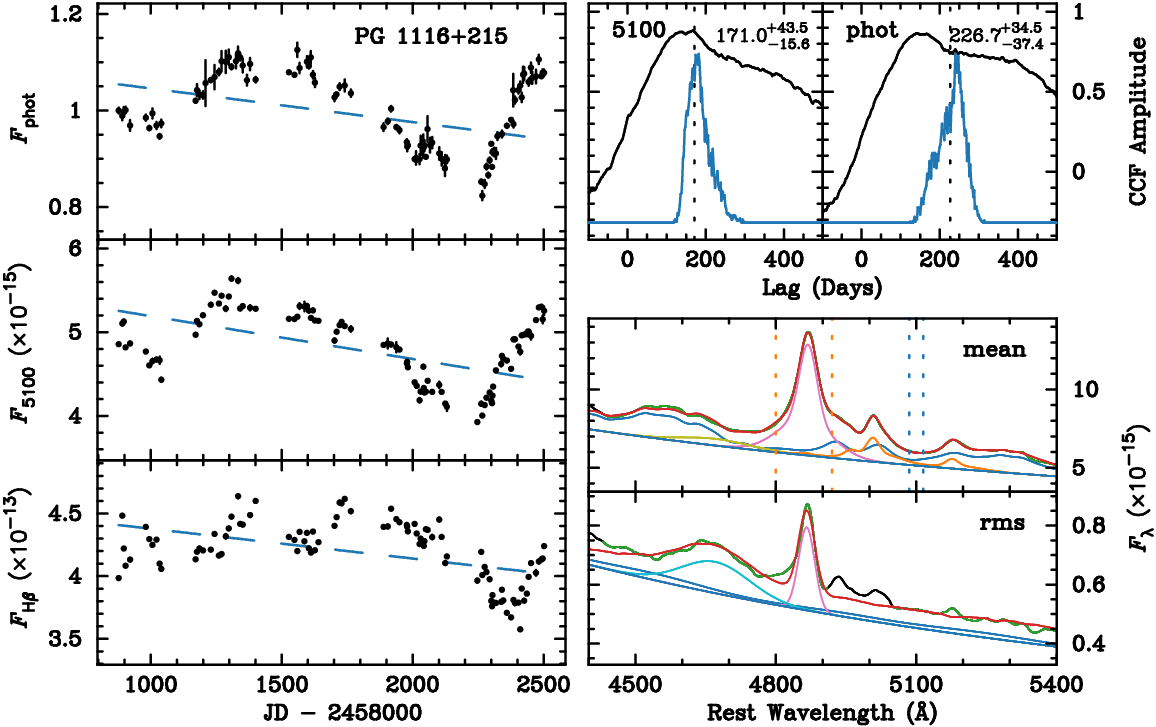}
  \caption{
  Light curves, ICCF analysis, and spectral decomposition of the mean and rms
  spectra for PG 1116+215, plotted in the same manner as in Figure
  \ref{fig-lc0157}. The blue dashed lines in the left column show the
  long-term trends of the light curves. See the text in Section \ref{sec-ccf}
  and Appendix \ref{sec-1116detrend} for the details, and Figure
  \ref{fig-1116detrend} for the results after detrending.
  }
  \label{fig-lc1116}
\end{figure*}

\begin{figure*}
  \centering
  \includegraphics[width=0.88\textwidth]{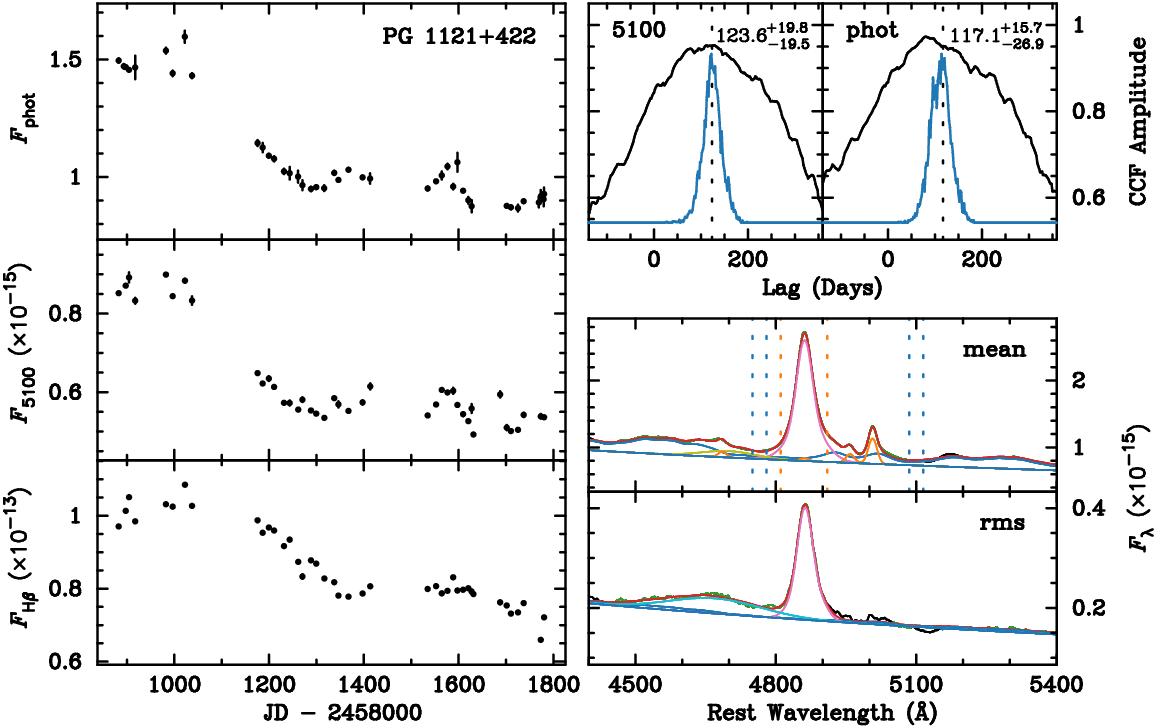}
  \caption{
  Light curves, ICCF analysis, and spectral decomposition of the mean and rms
  spectra for PG 1121+422, plotted in the same manner as in Figure
  \ref{fig-lc0157}.
  }
  \label{fig-lc1121}
\end{figure*}

\begin{figure*}
  \centering
  \includegraphics[width=0.88\textwidth]{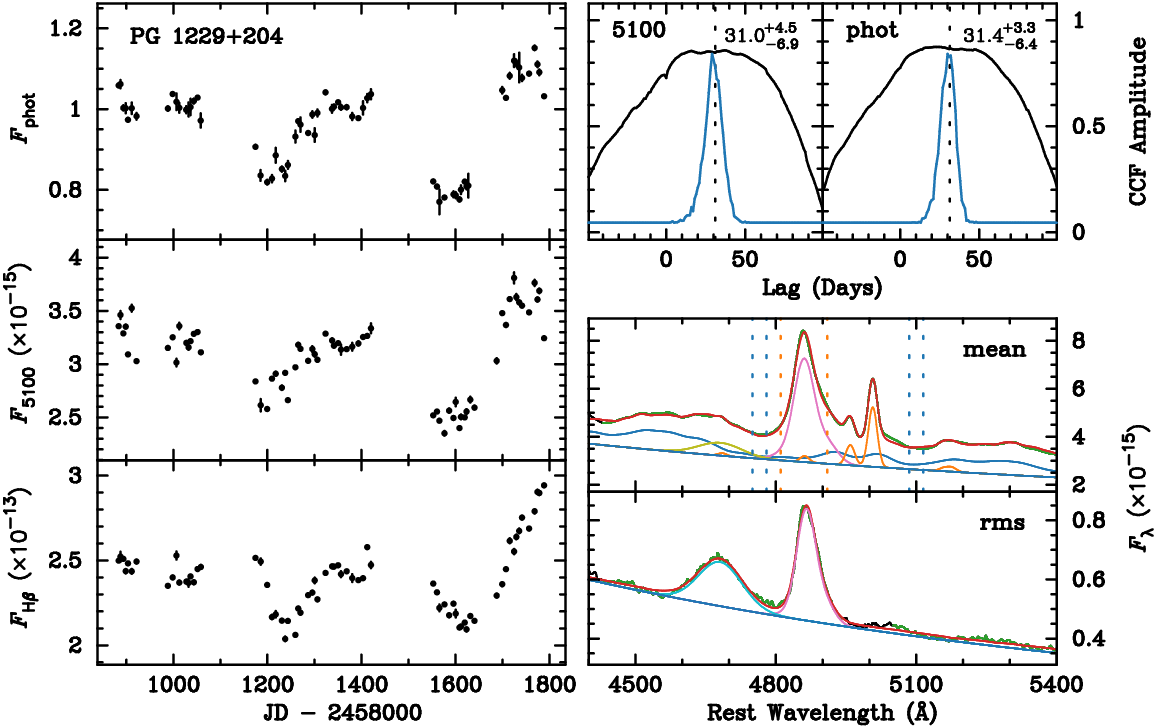}
  \caption{
  Light curves, ICCF analysis, and spectral decomposition of the mean and rms
  spectra for PG 1229+204, plotted in the same manner as in Figure
  \ref{fig-lc0157}.
  }
  \label{fig-lc1229}
\end{figure*}

\begin{figure*}
  \centering
  \includegraphics[width=0.88\textwidth]{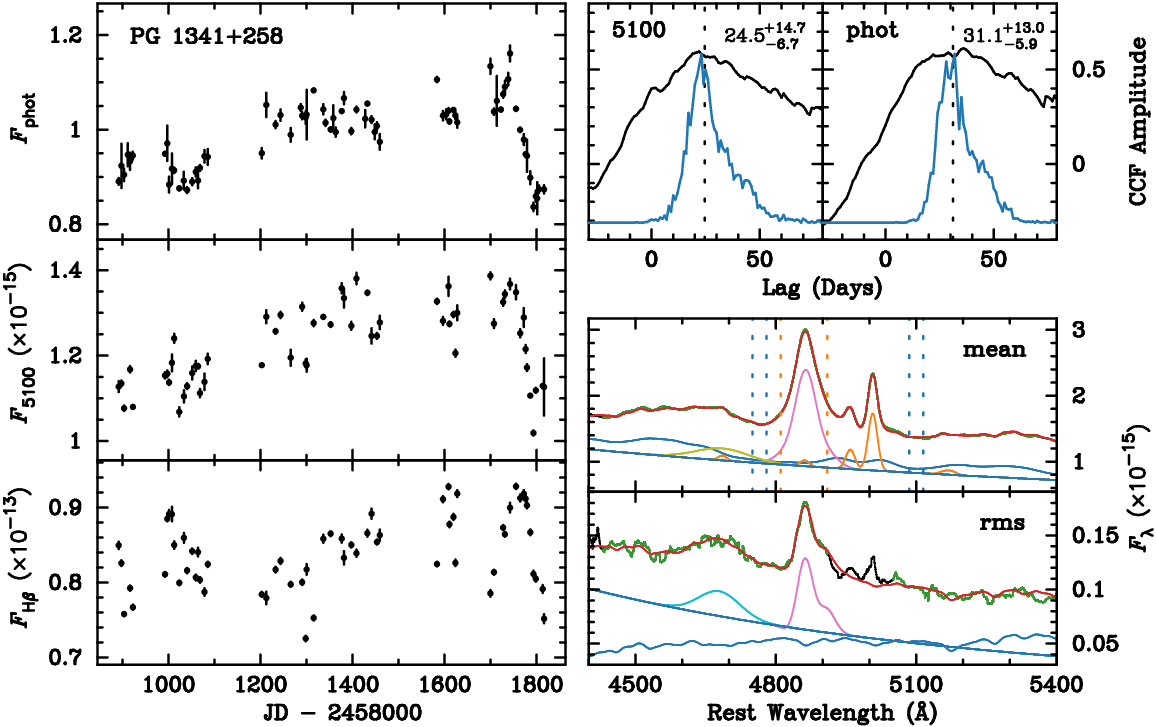}
  \caption{
  Light curves, ICCF analysis, and spectral decomposition of the mean and rms
  spectra for PG 1341+258, plotted in the same manner as in Figure
  \ref{fig-lc0157}.
  }
  \label{fig-lc1341}
\end{figure*}

\begin{figure*}
  \centering
  \includegraphics[width=0.88\textwidth]{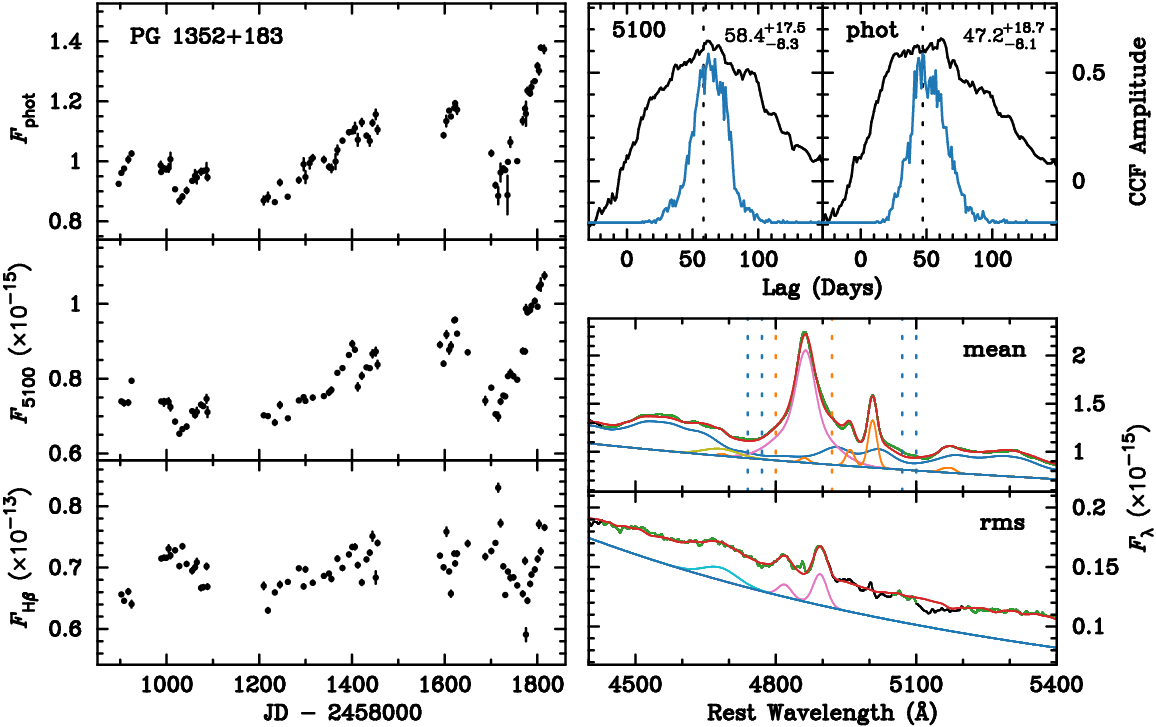}
  \caption{
  Light curves, ICCF analysis, and spectral decomposition of the mean and rms
  spectra for PG 1352+183, plotted in the same manner as in Figure
  \ref{fig-lc0157}.
  }
  \label{fig-lc1352}
\end{figure*}

\begin{figure*}
  \centering
  \includegraphics[width=0.88\textwidth]{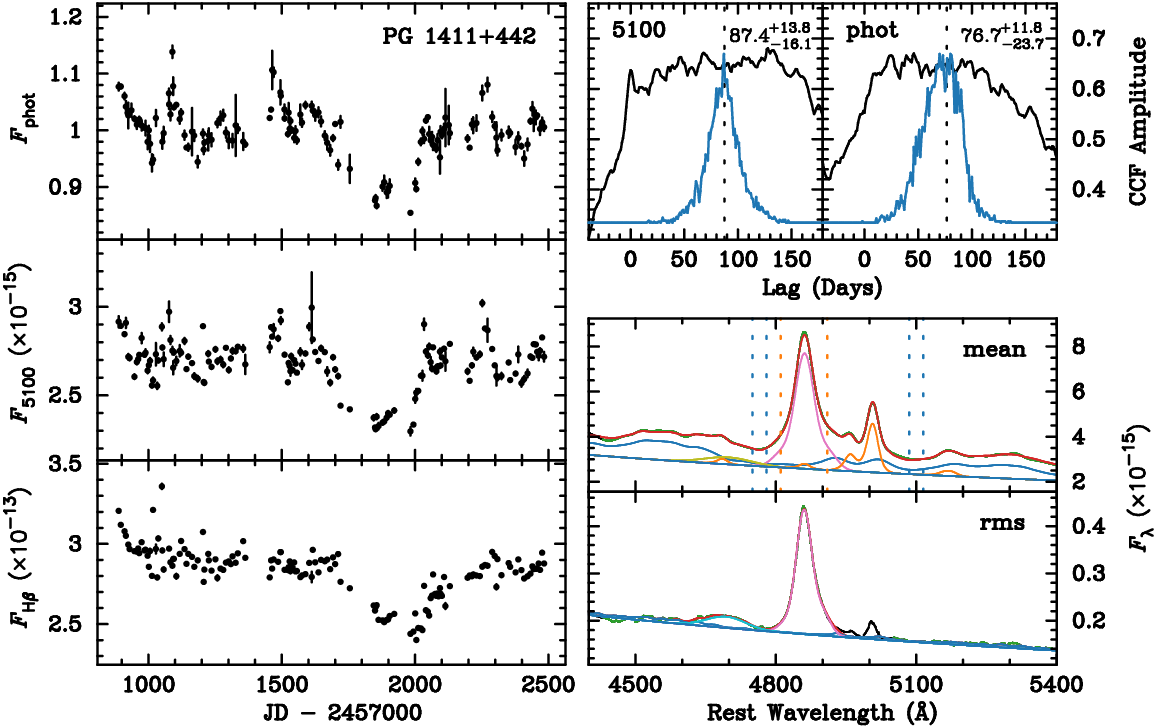}
  \caption{
  Light curves, ICCF analysis, and spectral decomposition of the mean and rms
  spectra for PG 1411+442, plotted in the same manner as in Figure
  \ref{fig-lc0157}.
  }
  \label{fig-lc1411}
\end{figure*}

\begin{figure*}
  \centering
  \includegraphics[width=0.88\textwidth]{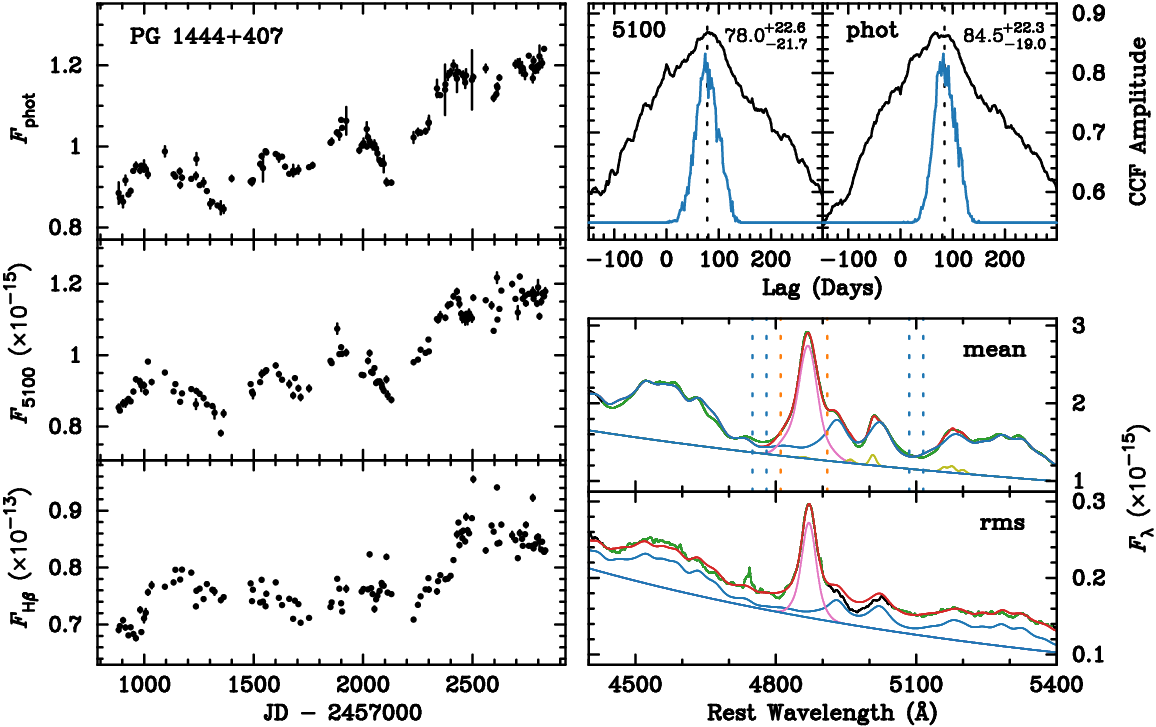}
  \caption{
  Light curves, ICCF analysis, and spectral decomposition of the mean and rms
  spectra for PG 1444+407, plotted in the same manner as in Figure
  \ref{fig-lc0157}.
  }
  \label{fig-lc1444}
\end{figure*}

\begin{figure*}
  \centering
  \includegraphics[width=0.88\textwidth]{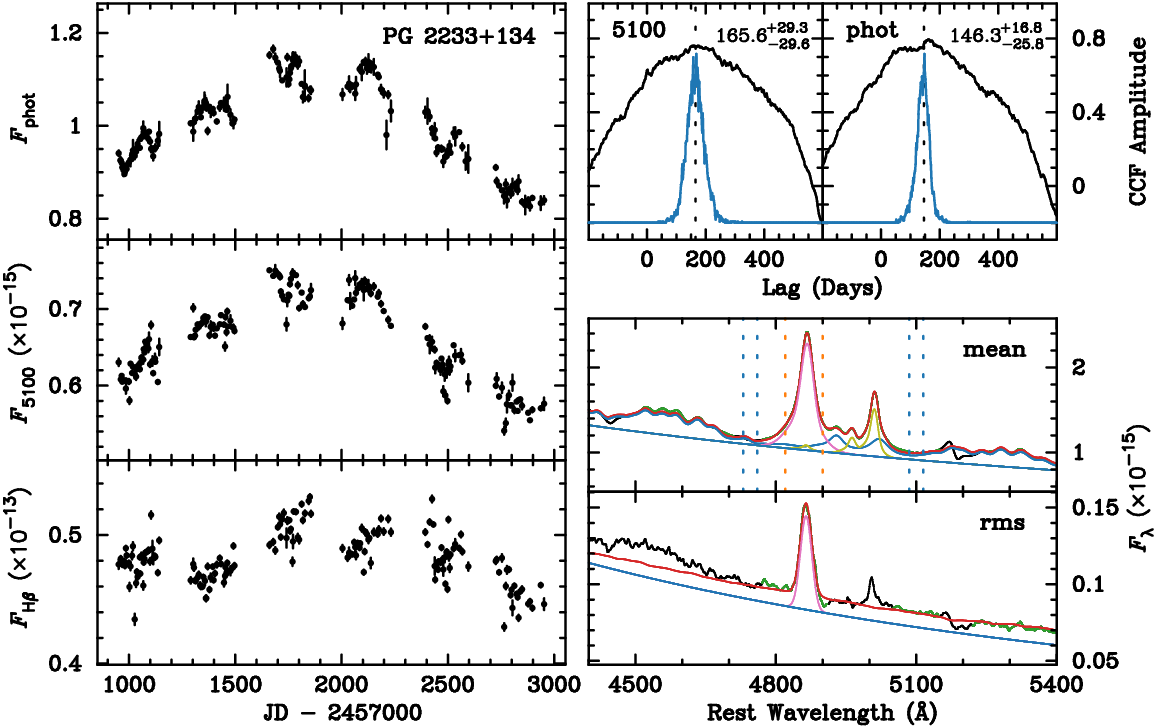}
  \caption{
  Light curves, ICCF analysis, and spectral decomposition of the mean and rms
  spectra for PG 2233+134, plotted in the same manner as in Figure
  \ref{fig-lc0157}.
  }
  \label{fig-lc2233}
\end{figure*}

\begin{figure*}
  \centering
  \includegraphics[width=0.88\textwidth]{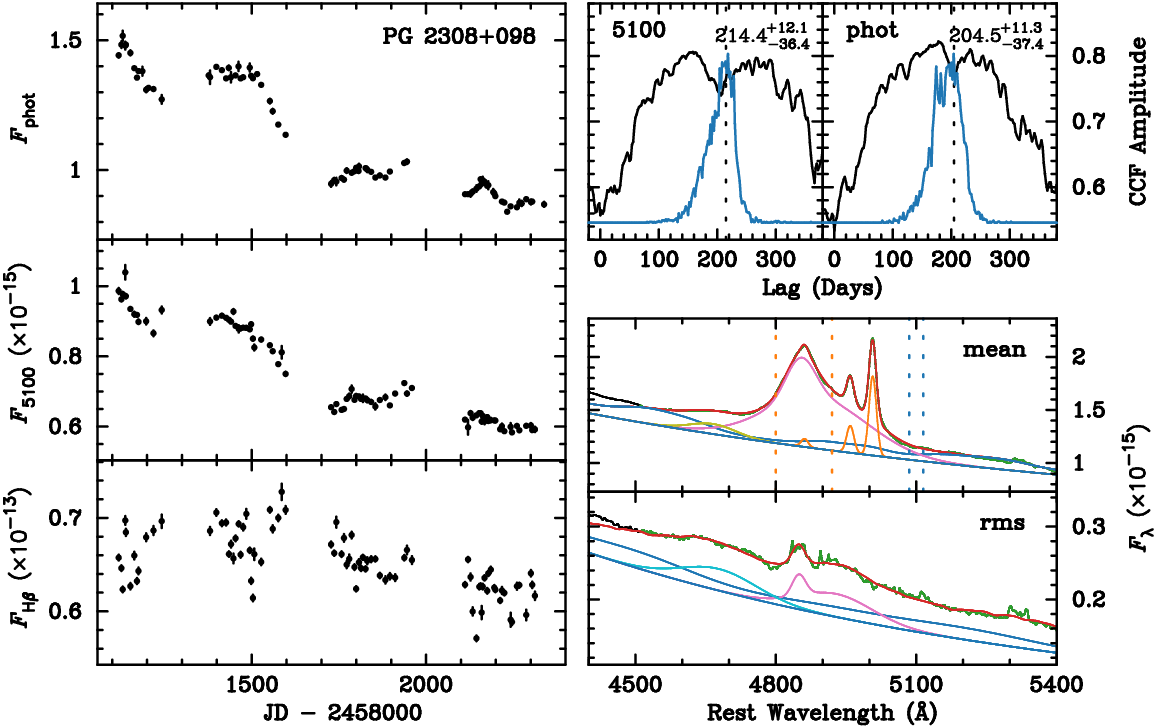}
  \caption{
  Light curves, ICCF analysis, and spectral decomposition of the mean and rms
  spectra for PG 2308+098, plotted in the same manner as in Figure
  \ref{fig-lc0157}.
  }
  \label{fig-lc2308}
\end{figure*}

We also corrected the small differences in the instrumental broadening and
wavelength shifting between different spectra, before generating the mean and
the root-mean-square residual (rms) spectra for each object (shown in the
lower two panels of the right column in each of Figures
\ref{fig-lc0157}--\ref{fig-lc2308}; note that these are corrected to the
rest-frame wavelength and for Galactic extinction). These differences could be
caused by the varying seeing (less than the 3$\farcs$0 slit width most of the
time) or a drift away from the center of the slit, and were corrected by
convolving each spectrum with a corresponding Gaussian determined by measuring
the velocity width and shift of the \oiii\ lines (see \citealt{hu21} and also
\citealt{hu16} for the details).

For PG 2308+098, two telluric-absorption bands happen to be located close to
the wavelengths of \hb\ and 5100 \AA\ (see Figure \ref{fig-2308abs} in
Appendix \ref{sec-2308abs}). Thus, additionally, telluric-absorption
correction were preformed by also using the comparison star as the telluric
standard \citep[e.g.,][]{kaspi00,lu21} for this object. Details of the
correction are described in Appendix \ref{sec-2308abs}. 

\section{Light-curve Measurements}
\label{sec-lc}

Following \citet{hu21}, the simple integration method is used here for
measuring the fluxes of the continuum and the broad \hb\ line. Compared to
spectral fitting, integration performs well in most cases (see the simulations
in the Appendix of \citealt{hu21}). Two situations have to be considered: 1)
when the host-starlight contribution is strong and causes ``apparent flux
variation'' which is an observational effect \citep{hu15}; 2) when the broad
\heii\ line is strong and highly variable and influences the continuum
measurement on the blue side of \hb\ (see an example in \citealt{hu20a}, their
Figure 2). For all the 11 objects in this work, the contributions of host
starlight to the continuum flux at 5100 \AA\ are not strong (this effect is
most pronounced in PG 1341+258 with $\sim$36 \%; see the final two columns of
Table \ref{tab-mass} below). The influence of the apparent flux variation of
the host is negligible for all the objects in this work, as indicated by the
consistency between the spectral 5100 \AA\ continuum light curves and the
$V$-band light curves (see Figures \ref{fig-lc0157}--\ref{fig-lc2308}, the
middle and top panels in the left column). The influence of the \heii\ broad
emission line to the continuum at the blue side of \hb\ can be evaluated by
the strength of broad \heii\ line in the rms spectrum, and moving the blue
continuum window to the flux valley between \hg\ and \hd\ (where \feii\
emission is weak; see the window C in Figure 2 of \citealt{hu20a}) could be a
good substitute as shown in \citet{hu20a} (see their Figures 2 and 3).

In each panel of the mean spectrum of Figures
\ref{fig-lc0157}--\ref{fig-lc2308} (in the middle of the right column), the
vertical dotted blue lines mark the windows for measuring the continuum
beneath the \hb\ line. The red window is centered at 5100 \AA\ and used also
for measuring the continuum light curve. The blue window is set at the valley
between \hb\ and the emission bump around 4600 \AA\ associated with \feii\ and
\heii\ for most of the objects except PG 1116+215 and PG 2308+098. The \heii\
lines of these two objects are very broad and highly variable (see their rms
spectra in Figures \ref{fig-lc1116} and \ref{fig-lc2308}, respectively), thus
their blue continuum windows are set to the valley between \hg\ and \hd\ as
mentioned above. For each object, after subtracting the straight line of the
continuum determined by the two continuum windows, the flux of \hb\ is
integrated in the line window (between the vertical dotted orange lines),
whose range is set to cover the most variable part of the emission line
indicated by its profile in the rms spectrum, except for in the case of PG
2308+098 (explained in the following paragraph).

For PG 2308+098, the \hb\ line in its rms spectrum shows a profile which is
complex and differs significantly from the profile in the mean spectrum (see
Figure \ref{fig-lc2308}, lower panels in the right column). The profile can be
modeled by a double-Gaussian, in which the narrow component is somewhat
blueshifted while the other, very-broad component has very large redward
shift. The broad shallow dip on the blue wing could be the residual of an
imperfect correction to the telluric absorption (see Section \ref{sec-obs} and
Appendix \ref{sec-2308abs}), while the red wing may have contributions from
the \feii\ emission and \oiii\ lines. In addition, the \heii\ line is highly
variable and very broad with its red wing extending beneath \hb. All these
effects make the choice of \hb\ integration windows for PG 2308+098 less clear
than for other objects. We tried several windows, and finally chose the
relatively narrow one which covers only the line core, shown by the orange
vertical dotted lines in the panel of the mean spectrum in Figure
\ref{fig-lc2308}: the blue limit is set to avoid the contribution from \heii,
while the red limit is set to exclude the \feii\ emission (mainly
$\lambda$4924 and $\lambda$5018) and the \oiii\ $\lambda\lambda$4959,~5007
lines. We found that this window yields the best \hb\ light curve (by means of
having high \rmax\ in the following ICCF analysis, see Section \ref{sec-ccf}
below) among our attempts. Note that it is possible that the \hb\ time lag of
PG 2308+098 given by this light curve is biased towards a longer value,
because the wings of the line, which is supposed to be emitted from the
high-velocity clouds, are not included in the integration of the line flux.

For the $V$-band light curves, the differential instrumental magnitudes for
the object and the comparison star with respect to other stars nearby in the
field of view were obtained by aperture photometry. The $V$-band light curves
of our comparison stars confirm that all of them were non-varying during our
campaign and thus suitable to be used for the spectroscopic flux calibration
(described in Section \ref{sec-obs} above).

The three panels in the left column of each of Figures
\ref{fig-lc0157}--\ref{fig-lc2308} show the resultant light curves of the
$V$-band (\fph, top panel, plotted in arbitrary linear units), the
spectroscopic 5100 \AA\ continuum (\fc, middle panel), and the \hb\ line
(\fhb, bottom panel), for each object respectively. For all the objects, the
light curves of photometric \fph\ and spectroscopic \fc\ are highly
consistent, confirming the high accuracy of our spectroscopic flux calibration
by the comparison star strategy. The data of these light curves for all the 11
objects is presented online as a machine-readable table, and a few sample
lines are shown in Table \ref{tab-lc} for illustration. Note that in the
table, the $V$-band flux is in units of instrumental magnitudes, and \fc\ and
\fhb\ are in the observed frame and not corrected for the Galactic extinction.

\begin{deluxetable}{lccr@{~$\pm$~}l}
  \tablewidth{0pt}
  \tablecaption{Light Curves
  \label{tab-lc}}
  \tablehead{
  \colhead{Object} & \colhead{Measure} & \colhead{JD} &
  \multicolumn{2}{c}{Flux} 
  \\
  \colhead{(1)} & \colhead{(2)} & \colhead{(3)} &
  \multicolumn{2}{c}{(4)}
  } 
  \startdata
0157+001 &  $V$ & 2457963.620 & 2.270 & 0.003 \\
0157+001 &  $V$ & 2457971.629 & 2.273 & 0.013 \\
\multicolumn{5}{c}{$\vdots$} \\
0157+001 & 5100 & 2457963.651 & 0.844 & 0.006 \\
0157+001 & 5100 & 2457971.653 & 0.839 & 0.007 \\
\multicolumn{5}{c}{$\vdots$} \\
0157+001 &  \hb & 2457963.651 & 0.328 & 0.002 \\
0157+001 &  \hb & 2457971.653 & 0.318 & 0.002 \\
\multicolumn{5}{c}{$\vdots$} \\
0844+349 &  $V$ & 2458393.683 & 0.468 & 0.011 \\
0844+349 &  $V$ & 2458397.646 & 0.484 & 0.011 \\
\multicolumn{5}{c}{$\vdots$} \\
0844+349 & 5100 & 2458393.700 & 7.031 & 0.086 \\
0844+349 & 5100 & 2458397.657 & 7.001 & 0.019 \\
\multicolumn{5}{c}{$\vdots$} \\
0844+349 &  \hb & 2458393.700 & 3.862 & 0.009 \\
0844+349 &  \hb & 2458397.657 & 3.960 & 0.011 \\
\multicolumn{5}{c}{$\vdots$}
  \enddata
  \tablecomments{
  Example lines of the online machine-readable table of the $V$-band,
  spectroscopic 5100 \AA\ continuum, and the \hb\ line light curves for the 11
  objects. Columns (1) and (2) shows the PG designation of the object and the
  name of the flux measurement, respectively. Columns (3) lists the Julia Date
  (JD). Column (4) lists the flux and the corresponding statistical
  uncertainty, in units of instrumental magnitudes, $10^{-15}$ \ergscma, and
  $10^{-13}$ \ergscm\ for the $V$-band, 5100 \AA\ continuum, and \hb,
  respectively.
  }
\end{deluxetable}

\section{Time-series Measurements}
\label{sec-ts}

For the spectroscopic light curves of 5100 \AA\ continuum and \hb\ plotted in
Figures \ref{fig-lc0157}--\ref{fig-lc2308} and listed in Table \ref{tab-lc},
the errors are only statistical and originate from just the uncertainties in
the observed counts of the spectra. Some systematic errors make the light
curves scatter above this level, introduced by, e.g., the flux-calibration
procedure, unstable slit losses, and host starlight contamination. Thus, for
each spectroscopic light curve, we used the same empirical method as in
\citet{hu21} to estimate an additional systematic error from the differences
in the fluxes of successive epochs. In Table \ref{tab-var}, we list the means
and the standard deviations for the fluxes of \fc\ and \fhb\ in columns (3)
and (6) respectively, and also the estimated systematic errors in the same
units in columns (4) and (7) correspondingly. These systematic errors had then
been added in quadrature before performing the following time-series analysis.

\begin{deluxetable*}{lr@{~$\pm$~}lr@{~$\pm$~}lcr@{~$\pm$~}lcr@{~$\pm$~}lcr@{~$\pm$~}l}
  \tablewidth{0pt}
  \tablecaption{Light curve statistics
  \label{tab-var}}
  \tablehead{
  \colhead{Object} & \multicolumn{2}{c}{$V$} &
  \multicolumn{5}{c}{5100 \AA} & &
  \multicolumn{5}{c}{\hb}
  \\ \cline{4-8} \cline{10-14}
  \colhead{} & \multicolumn{2}{c}{\fvar} &
  \multicolumn{2}{c}{Flux} & \colhead{$\sigma_{\rm sys}$} &
  \multicolumn{2}{c}{\fvar} & &
  \multicolumn{2}{c}{Flux} & \colhead{$\sigma_{\rm sys}$} &
  \multicolumn{2}{c}{\fvar}
  \\ \cline{4-6} \cline{10-12}
  \colhead{} & \multicolumn{2}{c}{(\%)} &
  \multicolumn{3}{c}{($10^{-15}$ \ergscma)} & \multicolumn{2}{c}{(\%)} & &
  \multicolumn{3}{c}{($10^{-13}$ \ergscm)} & \multicolumn{2}{c}{{(\%)}}
  \\
  \colhead{(1)} & \multicolumn{2}{c}{(2)} &
  \multicolumn{2}{c}{(3)} & \colhead{(4)} & \multicolumn{2}{c}{(5)} & &
  \multicolumn{2}{c}{(6)} & \colhead{(7)} & \multicolumn{2}{c}{(8)} 
  } 
  \startdata
PG 0157+001 & 14.6 & 0.8 & 1.063 & 0.151 & 0.016 & 14.1 & 0.8 & & 0.445 & 0.085 & 0.013 & 18.9 & 1.1 \\
PG 0844+349 & 10.5 & 0.6 & 6.631 & 0.672 & 0.110 & 10.0 & 0.6 & & 3.667 & 0.289 & 0.076 &  7.6 & 0.5 \\
PG 1116+215 &  7.7 & 0.6 & 4.829 & 0.419 & 0.070 &  8.5 & 0.6 & & 4.196 & 0.243 & 0.060 &  5.6 & 0.4 \\
PG 1121+422 & 20.3 & 2.3 & 0.624 & 0.125 & 0.013 & 19.9 & 2.2 & & 0.860 & 0.108 & 0.017 & 12.4 & 1.4 \\
PG 1229+204 & 10.6 & 0.9 & 3.086 & 0.368 & 0.078 & 11.6 & 1.0 & & 2.394 & 0.201 & 0.042 &  8.2 & 0.7 \\
PG 1341+258 &  7.3 & 0.6 & 1.224 & 0.092 & 0.031 &  7.0 & 0.7 & & 0.840 & 0.049 & 0.025 &  5.0 & 0.6 \\
PG 1352+183 & 11.8 & 1.0 & 0.809 & 0.104 & 0.016 & 12.7 & 1.1 & & 0.700 & 0.037 & 0.023 &  4.1 & 0.6 \\
PG 1411+442 &  4.4 & 0.3 & 2.683 & 0.143 & 0.047 &  4.9 & 0.3 & & 2.827 & 0.156 & 0.071 &  4.9 & 0.4 \\
PG 1444+407 & 11.1 & 0.7 & 1.009 & 0.117 & 0.017 & 11.4 & 0.8 & & 0.785 & 0.060 & 0.018 &  7.3 & 0.5 \\
PG 2233+134 &  9.0 & 0.5 & 0.660 & 0.054 & 0.010 &  8.0 & 0.5 & & 0.482 & 0.020 & 0.008 &  3.7 & 0.2 \\
PG 2308+098 & 19.5 & 1.5 & 0.744 & 0.132 & 0.011 & 17.7 & 1.4 & & 0.652 & 0.032 & 0.011 &  4.5 & 0.4
  \enddata
  \tablecomments{
  Variability amplitudes (\fvar) are listed in percentages. For spectroscopic
  5100 \AA\ continuum and the \hb\ line, the fluxes and the uncertainties
  listed are the means and standard deviations in the light curves. For each
  light curve, an estimated systematic error ($\sigma_{\rm sys}$) listed
  following the flux had been included in calculating the \fvar\ and
  estimating the uncertainty of the time lag later.
  }
\end{deluxetable*}

\subsection{Variability Amplitudes}

Columns (2), (5), and (8) of Table \ref{tab-var} list the variability
amplitudes (\fvar) for the three light curves, calculated according to the
definitions of \fvar\ and its uncertainty given by \citet{rodriguez97} and
\citet{edelson02}, respectively. \fvar\ represents the intrinsic variability
because both the statistical and systematic errors have been subtracted. The
\fvar\ of the $V$-band and 5100 \AA\ continuum light curves are consistent
with each other considering the uncertainties, indicating that our estimations
of the systematic errors in the spectroscopic light curves are reasonable. For
most of the objects (except PG 0157+001, see below), the \hb\ \fvar\ is
smaller than the continuum \fvar, which is commonly seen in reverberation
mapping observations and consistent with the results from photoionization
calculations \citep[e.g.,][]{korista04}.

PG 0157+001 is the only object in our sample that has a higher variability
amplitude in \hb\ than in the continuum. Note that the 5100 \AA\ continuum
flux is measured by integration and contaminated by the host starlight, thus
its variability amplitude could be underestimated. Taking the fraction of the
host contribution given by the spectral decomposition of the mean spectrum in
this object ($\sim$25\%; see Figure \ref{fig-lc0157} and Table \ref{tab-mass},
and the description in Section \ref{sec-width} below), the continuum
variability after correction is $\sim$19\%, close to that of \hb.

\subsection{Time Lags}
\label{sec-ccf}

The standard method of the interpolation cross-correlation function (ICCF;
\citealt{gaskell86,gaskell87,white94}) was employed to calculate the time lags
between the \hb\ and the continuum light curves. Following \citet{koratkar91}
and \citet{peterson04}, the centroid of the cross-correlation function (CCF)
above the 80\% level of the peak value (\rmax) was adopted as the measurement
of the lag, while the uncertainties were estimated according to the
cross-correlation centroid distribution (CCCD) generated by Monte Carlo
simulations via random subset selection (RSS) and flux randomization
\citep{maoz89,peterson98b}.

The resultant CCFs (black curves) and corresponding CCCDs (blue histograms)
are plotted in the two top-right panels in Figures
\ref{fig-lc0157}--\ref{fig-lc2308}. For each object, the left panel shows the
results for \hb\ with respect to the spectroscopic 5100 \AA\ continuum (\tsp),
while the right panel shows those for \hb\ with respect to the photometric
$V$-band light curve (\tph). The values of the lags, in the observed frame,
are marked as vertical dotted lines and also displayed with the uncertainties
as the numbers in corresponding panels. For summary, the \rmax\ and the \hb\
lags, in the rest frame, are listed in Table \ref{tab-ccf} for all the
objects. Columns (2) and (3) are the results for \hb\ with respect to the 5100
\AA\ continuum, while columns (4) and (5) are those for \hb\ with respect to
the $V$-band. Considering that the photometric $V$-band flux has more
contamination from the emission lines than the spectroscopic 5100 \AA\ flux,
especially for our sample in which the \feii\ emission is relatively strong,
\tsp\ is preferred to \tph\ in principle. Actually, for each object here, the
values of \tsp\ and \tph\ are very close, except in the case of PG 1116+215.

\begin{deluxetable}{lcr@{}lccr@{}l}
  \tablewidth{0pt}
  \tablecaption{Cross-Correlation Results
  \label{tab-ccf}}
  \tablehead{
  \colhead{Object} & \multicolumn{3}{c}{\hb\ vs. 5100 \AA} & &
  \multicolumn{3}{c}{\hb\ vs. $V$}
  \\ \cline{2-4} \cline{6-8}
  \colhead{} & \colhead{\rmax} & \multicolumn{2}{c}{Lag (\tsp)} & &
  \colhead{\rmax} & \multicolumn{2}{c}{Lag (\tph)}
  \\
  \colhead{} & \colhead{} & \multicolumn{2}{c}{(days)} & &
  \colhead{} & \multicolumn{2}{c}{(days)}
  \\
  \colhead{(1)} & \colhead{(2)} & \multicolumn{2}{c}{(3)} & &
  \colhead{(4)} & \multicolumn{2}{c}{(5)}
  } 
  \startdata
PG 0157+001 & 0.95 &  95.9 & $_{-11.4}^{+ 3.7}$ & & 0.96 &  74.0 & $_{ -6.6}^{+ 6.0}$ \\
PG 0844+349 & 0.85 &  38.5 & $_{ -3.7}^{+ 6.7}$ & & 0.86 &  43.0 & $_{ -3.9}^{+ 5.6}$ \\
PG 1116+215 & 0.88 & 145.5 & $_{-13.2}^{+37.0}$ & & 0.87 & 192.9 & $_{-31.8}^{+29.3}$ \\
PG 1121+422 & 0.95 & 100.9 & $_{-15.9}^{+16.2}$ & & 0.97 &  95.6 & $_{-22.0}^{+12.8}$ \\
PG 1229+204 & 0.86 &  29.1 & $_{ -6.5}^{+ 4.2}$ & & 0.87 &  29.5 & $_{ -6.0}^{+ 3.1}$ \\
PG 1341+258 & 0.59 &  22.6 & $_{ -6.1}^{+13.5}$ & & 0.61 &  28.6 & $_{ -5.4}^{+11.9}$ \\
PG 1352+183 & 0.65 &  50.7 & $_{ -7.2}^{+15.2}$ & & 0.66 &  41.0 & $_{ -7.0}^{+16.3}$ \\
PG 1411+442 & 0.68 &  80.2 & $_{-14.8}^{+12.7}$ & & 0.67 &  70.4 & $_{-21.7}^{+10.8}$ \\
PG 1444+407 & 0.87 &  61.6 & $_{-17.2}^{+17.9}$ & & 0.87 &  66.7 & $_{-15.0}^{+17.6}$ \\
PG 2233+134 & 0.76 & 125.0 & $_{-22.3}^{+22.1}$ & & 0.79 & 110.4 & $_{-19.5}^{+12.7}$ \\
PG 2308+098 & 0.81 & 149.6 & $_{-25.4}^{+ 8.5}$ & & 0.82 & 142.7 & $_{-26.1}^{+ 7.9}$
  \enddata
  \tablecomments{
  Peak values (\rmax) of the cross-correlation functions and the centroid time
  lags in units of days, in the rest frame. The lags of \hb\ vs. 5100 \AA\ are
  preferred and adopted for the following calculations.
  }
\end{deluxetable}

For PG 1116+215, the two lags have the largest difference in the relative
ratio in our sample: \tsp\ is $\sim$3/4 \tph, though they are still consistent
with each other considering the relatively large uncertainties. Both CCFs show
asymmetric shapes with rather slow declines to the long lags, making the
centroids deviate from the peaks (see Figure \ref{fig-lc1116}, top-right
panels). As shown in detail in Appendix \ref{sec-1116detrend}, this deviation
is caused by the different long-term trends in the light curves of this
object, which contribute to the long-lag extremities of the CCFs for both
\tsp\ and \tph\ but to different degrees. After detrending
\citep[e.g.,][]{welsh99} the light curves by subtracting a first-order
polynomial for each light curve (the blue dashed line in each panel of the
left column in Figure \ref{fig-lc1116}), the values of \tsp\ and \tph\ are
then nearly consistent with each other, and close to the value of \tsp\
without detrending. Considering the unknown origin of the long-term trend
here, we still adopt the lags without detrending, and \tsp\ is preferred.

For PG 2308+098, the light curves of the \hb\ and the continuum also seem to
have different long-term trends (see Figure \ref{fig-lc2308}), to an even more
severe degree than in the case of PG 1116+215 above. The continuum light
curves show just a long-term decline during the entire campaign, while the
\hb\ flux is increasing in the first year of our observation and with a much
flatter long-term trend. However, due to the long lag in this case, such a
difference in the long-term trends could be just an artifact. Thus, we
retrieved the light curve of this object from The All-Sky Automated Survey for
Supernovae (ASAS-SN; \citealt{shappee14,kochanek17}). Our monitoring
coincidentally happened to begin (marked by the vertical blue dashed line in
the top-left panel of Figure \ref{fig-2308as}) just after the flux peak in the
ASAS-SN light curve. By performing the ICCF analysis using the ASAS-SN light
curve as the continuum that extends $\sim$1.5 years earlier than our light
curves, the resultant \hb\ time lag is totally consistent with that resulting
from our spectroscopic continuum light curve, see Appendix \ref{sec-2308as}
for details. Thus, for uniformity, the value of time lag given by our light
curves is still adopted for the analysis below for this object.

For other objects showing long-term trends in their light curves, e.g., PG
1121+422 and PG 1444+407, their CCFs show no significant asymmetry.
Detrending their light curves has no actual impact on both \tsp\ and \tph,
considering the uncertainties in the measurements.

In summary, for all objects in our sample, \tsp\ calculated from the
spectroscopic \hb\ and 5100 \AA\ continuum light curves without detrending
(Table \ref{tab-ccf}, column 3) are adopted as the time lag measurements.

\section{Line Width Measurements}
\label{sec-width}

The widths of the broad \hb\ emission lines in both mean and rms spectra were
measured by the spectral fitting method the same as in \citet{hu21}. In brief,
after Galactic extinction correction (using the $A_V$ listed in Table
\ref{tab-obs}), the mean and rms spectra were decomposed into the following
spectral components on demand: the AGN continuum (a single power law), broad
\hb\ emission line (a double-Gaussian or Gauss-Hermite function), \feii\
emission (modeled from the \citealt{boroson92} template), the narrow emission
lines (a set of Gaussians), the host starlight (modeled from a
\citealt{bruzual03} simple stellar population template), and the broad \heii\
emission line (a Gaussian). See \citet{hu21} for more details of the modeling
of each component and the fitting.

The best-fit decompositions are shown in the two lower panels in the right
column of each of Figures \ref{fig-lc0157}--\ref{fig-lc2308}, for mean and rms
spectra of each object respectively. Note that the host starlight component is
often out of the panel view due to its low flux, but its intensity can be seen
from the departure between the total model and the power-law continuum plus
\feii\ emission. The host starlight component also appears in the rms
spectra of some objects, especially those with a strong host starlight
contribution (e.g., PG 1341+258). As described in \citet{hu15} and also the
Appendix of \citet{hu21}, this kind of apparent flux variation of host
starlight is due to the different slit losses for the extended host and the
point-like comparison star used for the flux calibration. Thus, the derived
flux of the host starlight could be overestimated by a changing factor due to
varying seeing and inexact slit centering. This effect contributes to the
systematic uncertainties in measuring the integrated \fc.

Both FWHM and the line dispersion (\sline) were calculated from the best-fit
model of the broad \hb\ component. For the rms spectra, the uncertainties in
the FWHM and \sline\ were estimated from the distributions of those measured
from the realizations generated by the RSS simulations (as in obtaining the
CCCDs). For the mean spectra, the uncertainties given by the RSS simulations
are much smaller than those introduced by the degeneracy of the narrow and the
broad \hb\ components (see also \citealt{hu15}). Thus, the differences in the
measurements of the broad \hb\ component in two fits of changing the flux
ratio of the narrow \hb\ component to \oiii\ $\lambda$5007 from 0\% to 20\%
are adopted as the uncertainties. The results are listed in columns (2)--(5)
of Table \ref{tab-mass}, the instrumental broadening and the extra broadening
by the convolution before generating the mean and the rms spectra (see Section
\ref{sec-obs}) have been corrected. Following are some notes on the objects PG
0157+001, PG 1352+183, and PG 2308+098.

\begin{deluxetable*}{lr@{~$\pm$~}lr@{~$\pm$~}lr@{~$\pm$~}lr@{~$\pm$~}lr@{}lr@{}lcr@{~$\pm$~}l}
  \tablewidth{0pt}
  \tablecaption{Line Widths, Virial Masses, and Luminosities
  \label{tab-mass}}
  \tablehead{
  \colhead{Object} &
  \multicolumn{2}{c}{FWHM$_{\rm mean}$} &
  \multicolumn{2}{c}{$\sigma_{\rm line,~mean}$} &
  \multicolumn{2}{c}{FWHM$_{\rm rms}$} &
  \multicolumn{2}{c}{$\sigma_{\rm line,~rms}$} & 
  \multicolumn{2}{c}{Virial Product} &
  \multicolumn{2}{c}{\mbh} &
  \colhead{$\lambda L_{\rm \lambda,gal}$(5100 \AA)} &
  \multicolumn{2}{c}{$\lambda L_{\rm \lambda,AGN}$(5100 \AA)}
  \\
  \colhead{} & 
  \multicolumn{2}{c}{(\kms)} &
  \multicolumn{2}{c}{(\kms)} & 
  \multicolumn{2}{c}{(\kms)} &
  \multicolumn{2}{c}{(\kms)} & 
  \multicolumn{2}{c}{($\times 10^7 M_\odot$)} &
  \multicolumn{2}{c}{($\times 10^7 M_\odot$)} &
  \colhead{($\times 10^{44}$ \ergs)} &
  \multicolumn{2}{c}{($\times 10^{44}$ \ergs)}
  \\
  \colhead{(1)} & 
  \multicolumn{2}{c}{(2)} &
  \multicolumn{2}{c}{(3)} & 
  \multicolumn{2}{c}{(4)} &
  \multicolumn{2}{c}{(5)} & 
  \multicolumn{2}{c}{(6)} &
  \multicolumn{2}{c}{(7)} & 
  \colhead{(8)} & 
  \multicolumn{2}{c}{(9)} 
  } 
  \startdata
PG 0157+001 &  2819 &  935 &  2077 & 429 &  1818 &   45 &  1220 &   38 &   2.78 & $_{ -0.37}^{+ 0.20}$ &  12.00 & $_{ -1.61}^{+ 0.88}$ &  1.12 &  3.18 &  0.65 \\
PG 0844+349 &  2497 &   35 &  1502 &  10 &  1905 &   62 &  1201 &   83 &   1.08 & $_{ -0.18}^{+ 0.24}$ &   4.67 & $_{ -0.79}^{+ 1.04}$ &  0.03 &  3.50 &  0.38 \\
PG 1116+215 &  3112 &   17 &  2728 &   7 &  2028 &  165 &   861 &   70 &   2.11 & $_{ -0.39}^{+ 0.64}$ &   9.08 & $_{ -1.69}^{+ 2.74}$ &  1.21 & 21.27 &  2.15 \\
PG 1121+422 &  2531 &   50 &  1661 &  13 &  2351 &   53 &  1294 &   44 &   3.30 & $_{ -0.57}^{+ 0.58}$ &  14.21 & $_{ -2.44}^{+ 2.48}$ &  0.17 &  5.26 &  1.16 \\
PG 1229+204 &  3284 &  155 &  1884 &  16 &  2998 &   88 &  1541 &  127 &   1.35 & $_{ -0.37}^{+ 0.30}$ &   5.82 & $_{ -1.61}^{+ 1.27}$ &  0.34 &  1.25 &  0.20 \\
PG 1341+258 &  3314 &  161 &  1838 &  21 &  2263 &  268 &  1588 &  386 &   1.11 & $_{ -0.62}^{+ 0.86}$ &   4.79 & $_{ -2.67}^{+ 3.70}$ &  0.43 &  0.75 &  0.09 \\
PG 1352+183 &  3412 &  213 &  2798 & 136 &  1730 & 1651 &  2223 &  648 &   4.89 & $_{ -2.93}^{+ 3.20}$ &  21.08 & $_{-12.65}^{+13.81}$ &  0.17 &  2.38 &  0.37 \\
PG 1411+442 &  2766 &   91 &  1847 &  34 &  2171 &   79 &  1401 &   57 &   3.07 & $_{ -0.62}^{+ 0.55}$ &  13.24 & $_{ -2.67}^{+ 2.35}$ &  0.43 &  2.25 &  0.15 \\
PG 1444+407 &  2766 &   17 &  1661 &   3 &  1914 &   62 &  1054 &   43 &   1.33 & $_{ -0.39}^{+ 0.40}$ &   5.75 & $_{ -1.67}^{+ 1.73}$ &  0.00 & 12.13 &  1.61 \\
PG 2233+134 &  1868 &   55 &  1523 &  23 &  1196 &   97 &   508 &   41 &   0.63 & $_{ -0.15}^{+ 0.15}$ &   2.71 & $_{ -0.66}^{+ 0.65}$ &  0.44 & 15.27 &  1.36 \\
PG 2308+098 &  9053 &  732 &  6131 & 135 &  9669 &  957 &  4780 &  492 &  66.72 & $_{-17.80}^{+14.25}$ & 287.58 & $_{-76.72}^{+61.43}$ &  0.00 & 33.57 &  6.71
  \enddata
  \tablecomments{
  The four measures of widths of the broad \hb\ line listed in columns
  (2)--(5) have been corrected for the instrumental broadening. The Virial
  Products listed in column (6) are calculated using the line dispersion
  \sline\ in the rms spectra, and then the masses of the black holes (column
  7) are estimated assuming a virial factor of 4.31 given by \citet{grier13}.
  The luminosities of the AGN continuum and the host starlight at 5100 \AA\
  are calculated from their fluxes given by the spectral decomposition to the
  mean spectra, while the uncertainty of the AGN luminosity is derived from
  the standard deviation of the 5100 \AA\ continuum flux during our campaign.
  }
\end{deluxetable*}

\textit{PG 0157+001.} In this object, the narrow emission lines are not only
rather strong (relative to the \hb), but also very broad (with a FWHM of 1019
\kms\ in the mean spectrum after correcting the instrumental broadening; the
two \oiii\ $\lambda\lambda$4959,5007 lines are even blended under the
resolution of our spectra, see Figure \ref{fig-lc0157}). Considering the
relatively low spectral resolution of our spectra, it is not easy to decompose
the narrow and the broad \hb\ components. Thus, our estimation of the
uncertainty in the width of the broad \hb\ line in the mean spectrum by
varying flux ratio of the narrow \hb\ to \oiii\ gives a rather large error,
especially for the FWHM ($\sim$30\%; see Table \ref{tab-mass}). On the other
hand, it is easy to measure the width of the broad \hb\ line in the rms
spectrum, because the narrow emission lines, including both \hb\ and \oiii,
disappear mostly. The relative uncertainties in the widths of \hb\ in the rms
spectra are just a few percentages, similar to those in other objects.

\textit{PG 1352+183.} The variability amplitude of \hb\ in this object is the
second weakest ($4.1\pm0.6$\%; Table \ref{tab-var}) among the 11 objects here.
Its \hb\ line in the rms spectrum shows a double-peaked profile, which can be
modeled by two separate Gaussians (see Figure \ref{fig-lc1352}). Thus, the
measurement of the \hb\ FWHM in the rms spectrum is highly uncertain: for most
of the realizations in the RSS simulations, only the red component is counted
for the peak of the blue component is lower than the half of that of the red
one. That's why the FWHM in the rms spectrum is even smaller than the \sline,
and the uncertainty of the FWHM is very large. On the other hand, the \sline\
measurement is more reliable, for both components are always included in the
calculations.

It is also worth noting here that such an \hb\ profile in the rms spectrum
indicates that during our campaign the variability of the \hb\ line in this
object happens mainly on the wings, which are supposed to be emitted by those
clouds with high velocities and close to the center. The core of the \hb\
line, which is supposed to be emitted from the outer part of the BLR, appears
to respond to only the long-term variability due to the so-called
``geometrical dilution'' effect \citep{goad14}. Considering that during our
campaign the continuum of this object shows a rather slow long-term
variability and a much fast short-term dip feature in the last year, it is
possible that the variability of the line core is smoothed out. The time lag
measured during our campaign may underestimate the size of the entire BLR.

\textit{PG 2308+098.} Similar to the case of PG 1352+183 above, the
variability amplitude of the \hb\ line in PG 2308+098 is also low
($4.5\pm$0.4\%; Table \ref{tab-var}). In addition, the profile of the broad
\hb\ line in the rms spectrum could be influenced by the imperfect
telluric-absorption correction and other emission lines (e.g., \heii\ and
\feii) and shows a peculiar shape, as described in Section \ref{sec-lc} above.
Thus, the measurement of \hb\ FWHM in the rms spectrum could be uncertain and
overestimated, which is the only one in this work whose \hb\ FWHM is broader
in the rms spectrum than in the mean spectrum.

The widths of the broad \hb\ emission lines are $\sim$0.1--0.2 dex narrower in
the rms spectra than in the mean spectra, measured in either FWHM or \sline.
This result is basically the same as that for the 15 objects in \citet{hu21}.
The line profile in the mean spectrum relates to the distribution of the line
luminosity of the BLR clouds, while the line profile in the rms spectrum
reflects the distribution of the responsivity of the BLR clouds. Thus the
result of narrower \hb\ in the rms spectrum than in the mean spectrum simply
supports the general idea of the change in \hb\ responsivity along the radius
given by the photoionization calculations: higher responsivity at larger
radius \citep[e.g.,][]{korista04,goad14}.

\section{Black Hole Masses}
\label{sec-mass}

Following \citet{peterson04}, we adopted the \sline\ of \hb\ in the rms
spectrum as the velocity width of the broad \hb\ emission line for estimating
the mass of the central black hole. The \hb\ line in the rms spectrum has the
advantage of representing the varying part that contributes in the time lag
measurements. For each object, the virial product (VP) was calculated as:
\begin{equation}
  {\rm VP} = \frac{c \tau_{\rm sp} \sigma_{\rm line,~rms}^2}{G}~,
\end{equation}
where $c$ is the speed of light, $G$ is the gravitational constant, \tsp\ is
the measured time lag between \hb\ and spectroscopic 5100 \AA\ continuum, and
$\sigma_{\rm line,~rms}$ is the velocity dispersion of the \hb\ emission line
in the rms spectrum. Then, the mass of the central black hole (\mbh) was
estimated as $M_{\rm BH} = f {\rm VP}$, assuming a dimensionless virial factor
$f$ \citep[e.g.,][]{peterson00} which represents all the unknown effects,
including e.g., geometry, kinematics, and the inclination angle of the BLR. In
principle, the value of $f$ should be a variable, and could be obtained for
each individual object through the dynamical modeling
\citep[e.g.,][]{pancoast11,li13}. In practice, an average value of $f$ is
often obtained by calibrations with mass measurements by other methods, e.g.,
the $M_{\rm BH}$--$\sigma_\ast$ relation \citep[e.g.,][]{onken04}, sometimes
considering the dependence on other properties, e.g., the bulge type of the
host galaxy \citep{ho14}. Here we simply adopt the averaged value of $f$=4.31
estimated by \citet{grier13}. Better estimations of the values of $f$ thus
more precise \mbh\ measurements are out of the scope of this work.

The results of VPs and \mbh\ are listed in columns (6) and (7) of Table
\ref{tab-mass}. The \mbh\ in these 11 quasars span a range of
$\sim$3--300$\times 10^7~M_\odot$, roughly a magnitude larger than the \mbh\
in the first batch of 15 quasars in \citet{hu21} ($\sim$0.5--20$\times
10^7~M_\odot$). Most objects in this work have a \mbh\ of $10 \pm 5 \times
10^7~M_\odot$.

Note that the widths measured in Section \ref{sec-width} refer to the entire
\hb\ profiles in the mean and the rms spectra. On the other hand, a portion
of the \hb\ fluxes on the wings with the highest velocities has not been
accounted for in the measurements of the \hb\ light curves in Section
\ref{sec-lc}, since the \hb\ profiles in the mean spectra are not fully
covered by the integration windows. Nevertheless, this mismatch will not
affect the \mbh\ estimated by the width of \hb\ in the rms spectrum, provided
that the \hb\ profiles in the rms spectrum are narrower than those in the mean
spectrum and most of the varying fluxes of \hb\ are captured. This requirement
holds for all our objects, with the exception of PG 2308+098. For PG 2308+098,
our calculated \mbh\ could be overestimated, due to potential misalignment in
line width and flux measurements resulting from its broad and complex \hb\
profile in the rms spectrum, as elaborated in Sections \ref{sec-lc} and
\ref{sec-width}.

From the best-fit model to the mean spectrum, we also derived the luminosities
of the host starlight ($\lambda L_{\rm \lambda,gal}$) and the AGN power-law
continuum ($\lambda L_{\rm \lambda,AGN}$) at 5100 \AA\ for each object, as
listed in columns (8) and (9) of Table \ref{tab-mass}. The redshifts listed in
column (3) of Table \ref{tab-obs} and cosmological parameters of $H_0=72~{\rm
km~s^{-1}}$ Mpc$^{-1}$, $\Omega_{m}=0.3$, and $\Omega_\Lambda=0.7$ were used
in the calculations.

\section{Discussion}
\label{sec-dis}

\subsection{Comparison with Previous Time Lag Measurements}

Eight objects in this work have their broad emission-line time lag published
for the first time. While the other three, PG 0844+349, PG 1229+204, and PG
1411+442, had been monitored and had their \hb\ time lag reported by
\citet{kaspi00}, but with observations of relatively low cadence. It has been
known that the measured time lag could be biased if the cadence is not high
enough \citep[e.g.,][]{grier08,hu21}. For PG 0844+349 the value of the
previous measured \hb\ time lag is doubtfully low, while for PG 1411+442 the
previous value seems rather long with large uncertainties. For PG 1229+204,
the previous measurement from the low sampling observations are also dubious
for its rather large uncertainties, although the value happens to be
consistent with the high-cadence measurements in this work.

\textit{PG 0844+349.} This object was observed by \citet{kaspi00} with a
relatively low cadence of only $\lesssim$50 epochs spread over seven years
(for comparison, in this work, there are 153 epochs in five years). Their
measured \hb\ time lag was $13_{-11}^{+14}$ days (in the observed frame),
given by a specifically defined CCF lower cut for calculating the centroid,
due to the noisiness of their data \citep{kaspi00}. The re-analysis by
\citet{peterson04} yielded a lag of $3.2_{-10.6}^{+13.2}$ days for \hb, marked
as uncertain. Both values are much shorter than their measurements of \ha\ and
\hg\ lags ($\sim$30--40 days), which also hints at the inaccuracy of the
measurements.  \citet{peterson04} suggested that such an inconsistency of lag
measurements from line to line was caused by inadequate time sampling. The
\hb\ lag of this target from our data ($41.0_{-3.9}^{+7.1}$ days, in the
observed frame) is more reasonable in a sense of being comparable with their
measurements of the \ha\ and \hg\ lags.

\textit{PG 1229+204.} The \hb\ time lag measured by \citet{kaspi00} had rather
large uncertainties ($36_{-18}^{+32}$ days, in the observed frame), while the
value given by the re-analysis of \citet{peterson04} was not improved much
($40.2_{-16.3}^{+29.4}$ days). The differences between the lags of the Balmer
lines are also large: the \ha\ lag was $\sim$2 times as large as the \hb\ lag,
while the lag of \hg\ was shorter than a half of that of \hb. There were only
33 epochs of spectroscopic observations in a span of seven years.
\citet{peterson04} suggested that the light curves of this object were also
undersampled and the lag measurements were dubious. In this work, our
measurement of the \hb\ lag of this object is obtained from 69 epochs of
spectroscopic observations in three years, which is adequate in sampling to
yield a value with much lower uncertainties ($31.0_{-6.9}^{+4.5}$ days, in the
observed frame) than before.

\textit{PG 1411+442.} The \hb\ lags measured by \citet{kaspi00} and
\citet{peterson04} also had large uncertainties ($118_{-71}^{+72}$ and
$135.4_{-67.2}^{+66.4}$ days, respectively, in the observed frame). As in the
case of PG 1229+204 above, their only 24 epochs of spectroscopic observations
were undersampled, thus making the measurements uncertain. Our data have 155
epochs. However, the variations of this object were rather weak during our
campaign, except for one slow-changing dip. The CCF shows a rather flat peak,
but the centroids are stable in the RSS/FR simulations (see Figure
\ref{fig-lc1411}). Thus, the uncertainties of the lag measured from our data
are still much smaller than those from the previous campaign. The value of the
lag from previous low-cadence observations is $\sim$1.5 times as long as that
from our high-cadence observations ($87.4_{-16.1}^{+13.8}$ days, in the
observed frame), which is consistent with an often-seen bias due to
undersampling \citep[see, e.g.,][]{hu21}.

Note that, in comparison to the values presented in Table 1 of
\citet{kaspi00}, the flux densities at 5100 \AA\ of the three objects in this
study (listed in Table \ref{tab-lc}, without subtracting the host starlight
and the \feii\ emission) exhibit changes of approximately $+$79\%, $+$44\%,
and $-$28\% for PG 0844+349, PG 1229+204, and PG 1411+442, respectively.
Assuming a relation of $R_{\rm BLR} \propto L^{0.5}$, the related changes in
the \hb\ time lag would be roughly $+$34\%, $+$20\%, and $-$15\%, which are
insufficient to explain the discrepancies between the earlier time-lag
measurements and the results in this study.

\subsection{CAHA PG quasars}

Combining with the 15 PG quasars already published in \citet{hu21}, we have
currently obtained reliable \hb\ time lag measurements of 26 PG quasars in
total. All the 26 objects show ``normal'' reverberation response between the
\hb\ line and the continuum emission. Another two objects, PG 2130+099
\citep{hu20b,yao24} and PG 0026+129 \citep{hu20a}, both have some special
properties of reverberation (i.e., exhibiting significant changes in the
measurement of \hb\ lag due to ``geometric dilution'' for PG 2130+099 and
having a very-broad \hb\ component with a nearly zero lag for PG 0026+129),
thus are not included in the discussion here.

\begin{figure}
  \centering
  \includegraphics[width=0.45\textwidth]{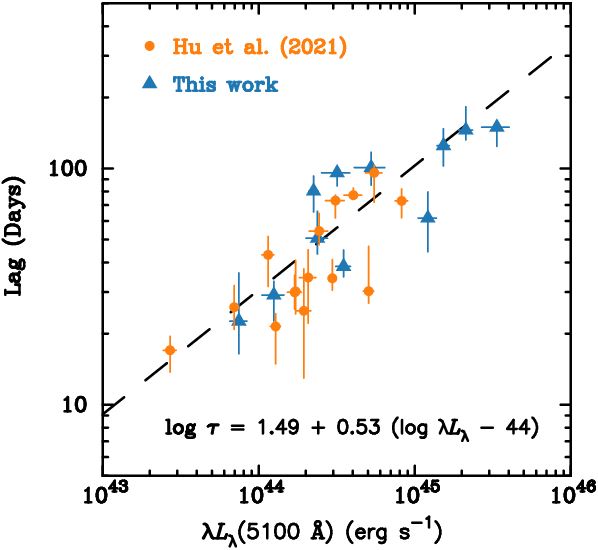}
  \caption{
  The \hb\ time lag versus AGN luminosity at 5100 \AA\ for the 26 objects in
  the current CAHA PG quasars reverberation mapping sample. The 11 objects in
  this work are marked as blue triangles, while the 15 objects previously
  published in \citet{hu21} are marked as orange cycles. The dashed line
  is the fit to the entire sample of 26 objects, whose expression is also
  displayed at the bottom.
  }
  \label{fig-sample}
\end{figure}

Figure \ref{fig-sample} shows the distribution of the current CAHA PG sample
on the diagram of \hb\ lag versus optical luminosity at 5100\AA. On average,
the 11 objects (blue triangles) in this work are $\sim$2.3 times as luminous
as the 15 objects (orange cycles) in \citet{hu21} ($4.9\times10^{44}$ versus
$2.1\times10^{44}$ \ergs), while the time lags are also $\sim$1.8 times as
long (69 versus 39 days). In total, the 26 targets span a rather wide range of
luminosity (more than 2 orders of magnitude), and it is suitable for studying
the \rl\ relation with our observations of homogeneous qualities of duration,
cadence, and flux calibration. The dashed line shows the fit to all of the 26
objects using the FITEXY method \citep{press92}:%
\begin{eqnarray}
  \log \left(\frac{R_{{\rm BLR}}}{\rm 1\,ltd}\right) = (1.49 \pm 0.03) +
  (0.53 \pm 0.04)\times
  \nonumber \\
  \log \left(\frac{\lambda L_{\lambda}}{10^{44}\,{\rm erg\,s^{-1}}}\right)~,
\end{eqnarray}
which is close to both the relation given by \citet{bentz13} and that for low
accretion rates AGNs in \citet{du18b}. This sample is slightly biased
towards high \rfe\ as shown in Section \ref{sec-samp} and Figure \ref{fig-ev}
but with limited impact on the \rl\ relation. Actually, the \rfe\ of most
objects in this sample are not as large as those of the objects with
significantly shortened time lags in \citet{du19} (see the top-left panel of
their Figure 4 and our Figure \ref{fig-ev}).

Note that the AGN optical luminosity here is given by the spectral fitting to
the mean spectrum. The precision of the decomposition of the host starlight
from the mean spectrum is constrained by the single template of simple stellar
population we used. The more precise measurement of the AGN continuum
luminosity requires removing the contribution of the host starlight using high
spatial-resolution imaging, which is beyond the scope of this paper. The
measurements of the lags of other emission lines, e.g., \feii\ and \heii,
require spectral fitting, and will be presented in a forthcoming paper.

\section{Summary}

We conducted a long-term reverberation mapping campaign since 2017 May at
Calar Alto Observatory, aiming to spectroscopically monitor PG quasars with
both high cadence and long duration. Here we present the results of our
observations until 2024 July, for the second batch of 11 PG quasars, including
the light curves of the photometric $V$-band, spectroscopic 5100 \AA\
continuum, the \hb\ broad line, their time-lag measurements, and estimations
of the mass of the central black holes, as summarized below.

\begin{enumerate}
  \item Reliable time lags are measured between the broad \hb\ emission line
    and the AGN continuum for 11 PG quasars. Our measurements of the \hb\ time
    lags are for the first time for eight objects. While for the other three,
    only uncertain \hb\ lag measurements exist in the literature from previous
    observations with relatively low quality time sampling.
  \item The widths of the broad \hb\ emission lines, the masses of the central
    black holes, and the AGN optical continuum luminosities are obtained from
    our observations. The black hole masses span a range of
    $\sim$3--300$\times 10^7~M_\odot$, while the AGN luminosities at 5100 \AA\
    range from $\sim$0.75 to 34 $\times 10^{44}$ \ergs, which are relatively
    high among objects with reverberation mapping measurements in the
    literature.
  \item Combining with the first batch of 15 PG quasars presented in
    \citet{hu21}, we have successfully monitored a sample of 26 PG quasars
    with uniformly high quality data and representative properties at Calar
    Alto Observatory. A tentative \rl\ relation with a slope of 0.53 between
    the BLR size and the AGN luminosity is obtained for these 26 objects.
\end{enumerate}

More analysis of this data set, including time-lag measurements for emission
lines other than \hb, velocity-resolved delays, dynamical modeling,
etc., will be presented in forthcoming papers.

\begin{acknowledgments}
We acknowledge the staff of the CAHA 2.2m telescope and others who support the
observations. This work is based on observations collected at the Centro
Astron\'omico Hispano en Andaluc\'ia (CAHA) at Calar Alto, operated jointly by
the Andalusian Universities and the Instituto de Astrof\'isica de Andaluc\'ia
(CSIC). This research is supported by the National Key R\&D Program of China
(2021YFA1600404 and 2023YFA1607904), by the National Science Foundation of
China (NSFC; 11833008, 11991050, 12122305, and 12333003). YRL acknowledges
financial support from the NSFC through grant No. 12273041 and from the Youth
Innovation Promotion Association CAS. PD acknowledges financial support from
NSFC grants NSFC-12022301 and 11991051. LCH acknowledges financial support
from the NSFC (11721303, 11991052, 12011540375, and 12233001), the National
Key R\&D Program of China (2022YFF0503401), and the China Manned Space Project
(CMS-CSST-2021-A04, CMS-CSST-2021-A06). JA acknowledges financial support from
the State Agency for Research of the Spanish MCIU through the ``Center of
Excellence Severo Ochoa'' award to the Instituto de Astrof\'isica de
Andaluc\'ia (SEV-2017-0709).
\end{acknowledgments}

\appendix

\section{Detrending in PG 1116+215}
\label{sec-1116detrend}

\begin{figure}
  \centering
  \includegraphics[width=0.45\textwidth]{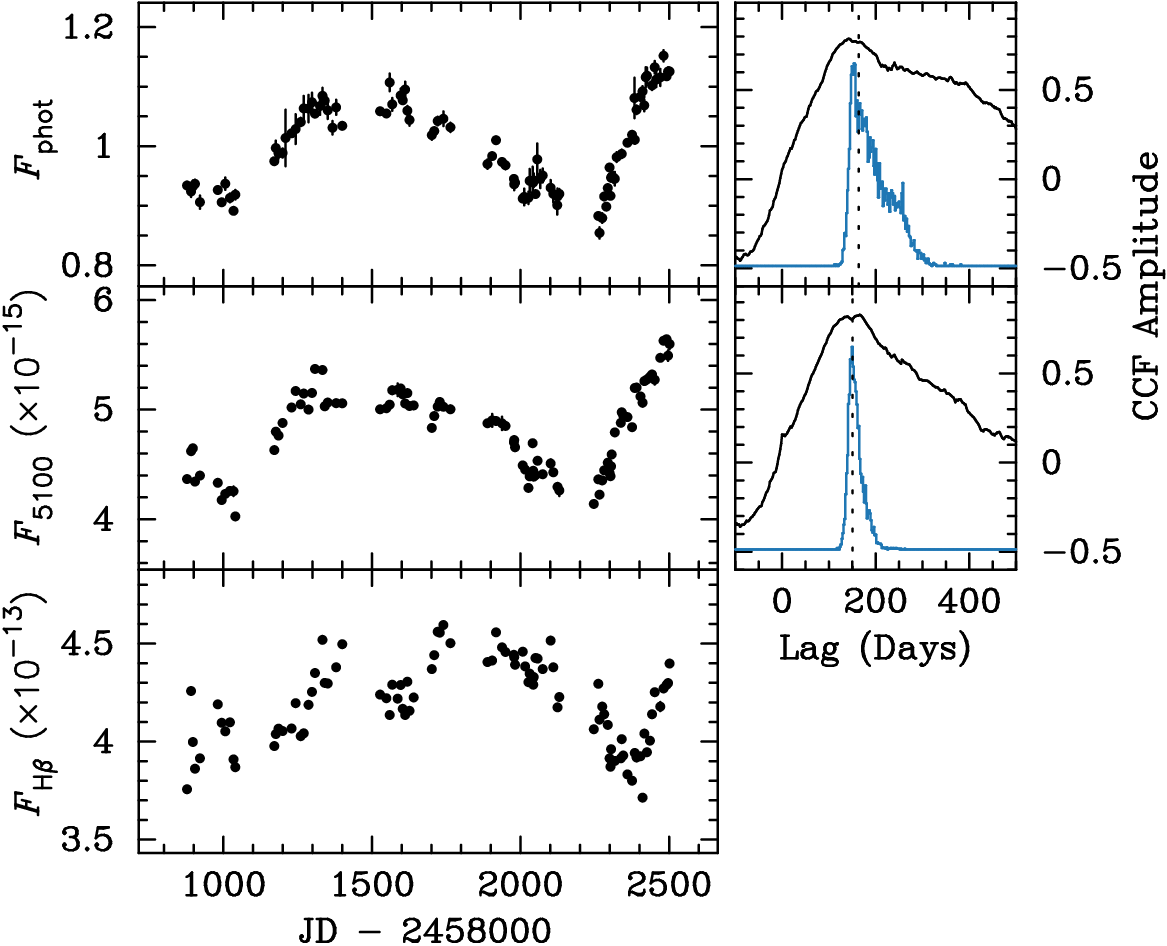}
  \caption{
  Detrended light curves (left column, from top to bottom: \fph, \fc, and
  \fhb) and corresponding ICCF results for \fhb\ with respect to \fph\ (right
  column, top panel) and \fc\ (right column, middle panel) for PG 1116+215.
  For each panel in the right column, the CCF and the corresponding CCCD are
  shown in black and blue, respectively, while the vertical dotted line marks
  the position of the centroid as the lag measurement. 
  }
  \label{fig-1116detrend}
\end{figure}

For comparison, we also applied ICCF analysis to the detrended light curves of
PG 1116+215, in which the long-term trends have been removed. The detrending
was preformed by subtracting the first-order polynomial fitted for each light
curve (as shown in the panels of the left column of Figure \ref{fig-lc1116},
blue dashed lines). The detrended $V$-band, 5100 \AA\ continuum, and \hb\ line
light curves are shown in the left column of Figure \ref{fig-1116detrend},
from top to bottom. And the ICCF results for \hb\ light curve with respect to
$V$-band and 5100 \AA\ continuum light curves are plotted in the right-top and
right-middle panels, respectively. After detrending, the CCFs (black curves)
become less asymmetric (see Figure \ref{fig-lc1116} for comparison),
especially that for \fhb\ with respect to \fc. Then the centroids (vertical
dotted lines) are more consistent with the peaks, and the CCCDs (blue
histograms) are peaked at the centroid although still asymmetric to have a
tail extended to the long lags. The values of measured lags are
$163.6_{-10.5}^{+70.7}$ days (\rmax = 0.79) for \fhb\ versus \fph, and
$150.7_{-4.9}^{+18.0}$ days (\rmax = 0.83) for \fhb\ versus \fc, both in the
observed frame. The two values are consistent with each other, and also
consistent with the value given by the original \hb\ and \fc\ continuum
without detrending ($171.0_{-15.6}^{+43.5}$ days, in the observed frame; see
the \fhb\ vs. \fc\ CCF panel of Figure \ref{fig-lc1116}).

\section{Notes for PG 2308+098}

\subsection{Telluric-absorption correction}
\label{sec-2308abs}

\begin{figure}
  \centering
  \includegraphics[width=0.45\textwidth]{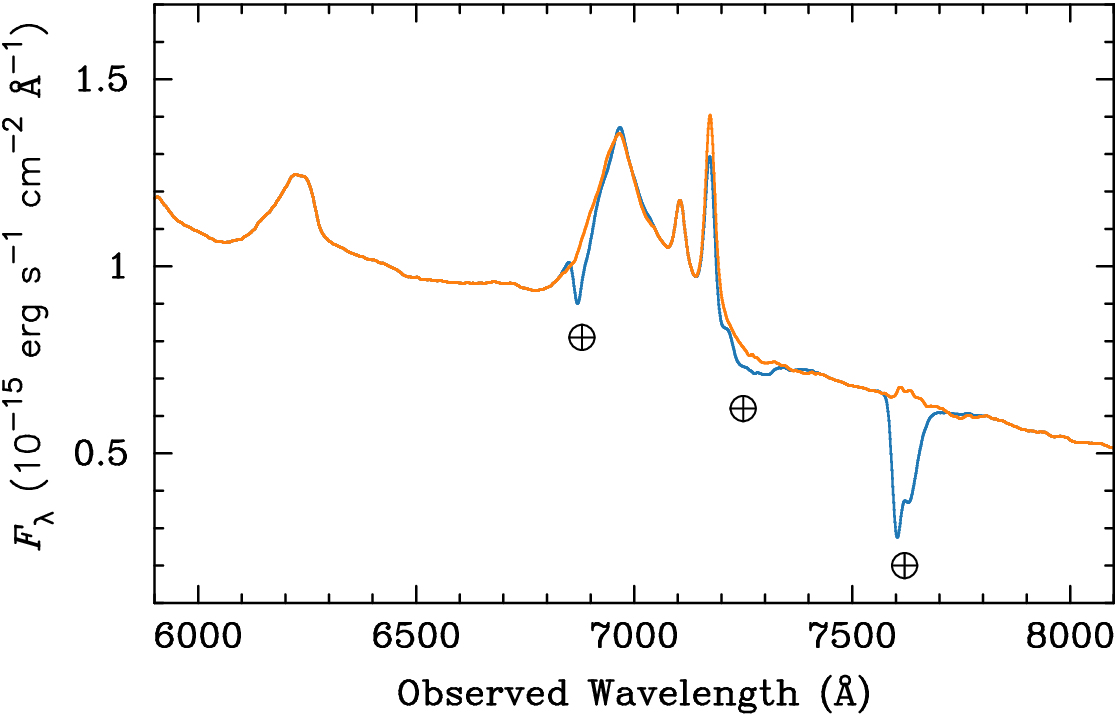}
  \caption{
  The mean spectra of PG 2308+098 before (in blue) and after (in orange) the
  telluric-absorption correction in three bands (marked by $\oplus$). 
  }
  \label{fig-2308abs}
\end{figure}

With a redshift of 0.4329, the highest of the 11 objects of this work, the
spectrum of PG 2308+098 around the \hb\ line is affected by several
telluric-absorption bands, primarily the blue wing of \hb\ and the continuum
close to 5100 \AA\ (see Figure \ref{fig-2308abs}; marked by $\oplus$). We
corrected the telluric absorption employing the simultaneously observed
comparison star which also served as a flux standard, similar to the methods
used in \citet{kaspi00} and \citet{lu21}. The blue and orange spectra in
Figure \ref{fig-2308abs} are the mean spectra before and after the
telluric-absorption correction, respectively. After correction, the blue wing
of \hb\ becomes rather smooth, and the flux ratio of \oiii\ $\lambda$5007 to
$\lambda$4959 is consistent to the theoretical value of 3.

\subsection{Continuum light curve from ASAS-SN}
\label{sec-2308as}

\begin{figure}
  \centering
  \includegraphics[width=0.45\textwidth]{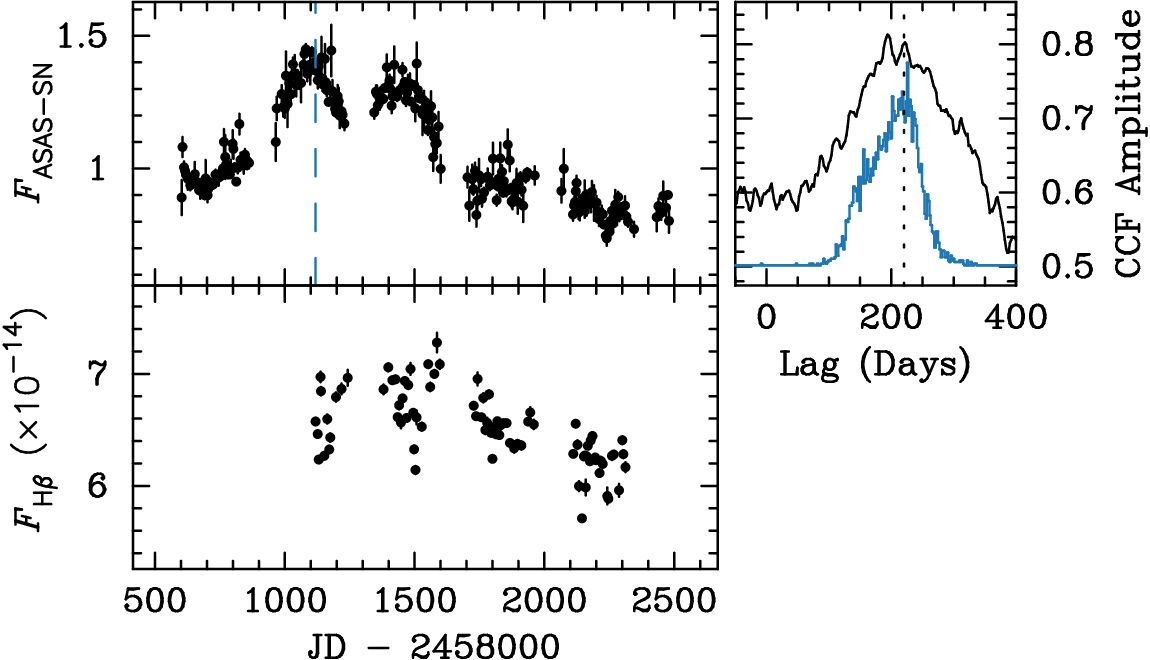}
  \caption{
  ICCF analysis for the continuum light curve from ASAS-SN (upper-left) and
  our \hb\ light curve (lower-left) for PG 2308+098. The vertical blue dashed
  line in the upper-left panel marks the beginning of our spectroscopic
  monitoring to this object. The upper-right panel shows the CCF (in black)
  and the corresponding CCCD (in blue). The vertical dotted line marks the
  centroid as the measurement of the time lag.
  }
  \label{fig-2308as}
\end{figure}

Considering the long lag ($\sim$200 days, in the observed frame) between the
\hb\ line and the continuum of PG 2308+098, it is valuable to extend the
continuum light curve to earlier times. We retrieved the light curve of this
object $\sim$1.5 years before the beginning of our spectroscopic monitoring
(marked by the vertical blue dashed line) from ASAS-SN
\citep{shappee14,kochanek17}, as shown in the upper-left panel of Figure
\ref{fig-2308as}. The data points with observation times closer than 3 days
have been averaged. ICCF analysis yields a lag of $220.4_{-63.2}^{+19.7}$ days
(in the observed frame) with a \rmax\ of 0.81 between the \hb\ light curve
measured from our spectra and the ASAS-SN light curve as the continuum. This
value is totally consistent with that obtained by our one-year-short
spectroscopic 5100 \AA\ continuum light curve ($214.4_{-36.4}^{+12.1}$ days,
in the observed frame; see the \fhb\ versus \fc\ CCF panel of Figure
\ref{fig-lc2308}).

\end{document}